\documentclass{aastex631}

\newcommand{\ms}{\mbox{m s$^{-1}~$}}

\newcommand{\kms}{\mbox{km s$^{-1}~$}}
\newcommand{\cm}{\mbox{cm$^{-1}~$}}
\newcommand{\cms}{\mbox{cm s$^{-1}~$}}

\newcommand{\degC}{$^\circ$C}

\submitjournal{AJ}

\graphicspath{{./}{figures/}}

\begin{document}

\title{Design and Commissioning of an Iodine Cell for the ESPRESSO Spectrograph}

\correspondingauthor{Gillian Nave}
\email{gilliannave@gmail.com}

\author[0000-0002-1718-9650]{Gillian Nave}
\affiliation{National Institute of Standards and Technology, Gaithersburg, 
MD 20889-8422, USA }
\affiliation{Physics Dept., Imperial College London, London SW7 2AZ, UK}

\author[0000-0003-1305-3761]{R. Paul Butler}
\affiliation{Carnegie Institution for Science, Earth \& Planets Laboratory, 
5241 Broad Branch Road NW, Washington, DC 20015-1305, USA}

\begin{abstract}

High resolution echelle spectrographs remain the backbone of precision Doppler
radial velocity (RV) programs, detecting almost all known exoplanets
within 50 pc. Precision Doppler RV spectrographs have traditionally fallen
into two camps. Standard unstabilized echelles observe targets through an iodine
absorption cell. The iodine spectrum is embedded on the target spectrum, and
provides a wavelength scale and a record of the spectrometer
point-spread-function (PSF). Super-stabilized spectrometers are placed inside a
vacuum tank, temperature stabilized at the level of 0.001~\degC, and fed by two
scrambled fibers. One fiber carries the target, the other the calibration source
(ThAr, Fabry-P\'{e}rot, laser-comb). Both techniques have found hundreds of
planets and produce sub \ms  uncertainties on the highest resolution
echelles currently available. Both techniques have advantages and disadvantages
and can be combined with the goal of reducing the long-term Doppler RV
uncertainty to the sub 10 \cms level.

We have designed, built, calibrated, and commissioned an iodine cell for the
European Southern Observatory's (ESO) ESPRESSO spectrograph. The design and
construction of the cell was carried out in 2022. The cell was calibrated at the
National Institute of Standards Technology (NIST) Atomic Spectroscopy laboratory
in early 2023 and was commissioned in May 2023. The commissioning run was
limited to evening and morning twilight on VLT-UT2. Five main sequence dwarf
stars ranging in spectral type from G to early K were observed between 4 and 6
nights spanning a total of ten nights.

\end{abstract}


\section{Introduction} \label{sec:intro}

From 1920 to 1980 Doppler radial velocity (RV) measurement 
uncertainty was stalled at the 300 \ms to 1 \kms level. In their seminal 1973
paper, \citet{Griffin_1987} outlined several of the systematic errors that limit
measurement uncertainty. Foremost among these is the wavelength
calibration. Historically the wavelength calibration has been carried out with
an emission lamp (e.g. Thorium-Argon) that is taken before and/or after the
stellar spectrum. The calibration is thus not taken simultaneously with the
stellar source. The calibration and stellar source also illuminate the entrance
slit and spectrometer optics differently. The lamp uniformly illuminates the
slit while starlight typically over-fills the slit and moves about the slit due
to guiding errors. Moving the centroid of the stellar image on the slit by a
one-tenth of a slit width results in a systematic Doppler offset relative to the
calibration of several hundred \ms. To overcome these problems, the Griffins
suggested observing the telluric O$_2$ band near 6300~\AA\ as a wavelength
standard. The advantage of this technique is that the reference spectrum
(telluric O$_2$) is carried on the beam on starlight, thus accounting for
guiding uncertainties and temporal variations in the spectrometer.

Starting in the 1980s several groups began experimenting with techniques to
improve Doppler RV uncertainty by two orders of magnitude, which would
allow for the detection of Jupiter and Saturn analogs around nearby stars. Among
the techniques explored were a fiber fed Fabry-P\'{e}rot transmission
interferometer calibrated with an Fe/Ar hollow cathode lamp (HCL)
\citep{McMillan_1986,Smith_1987}, a Fabry-P\'{e}rot in reflection
\citep{Cochran_1985}, and using the telluric O$_2$ band as a wavelength
reference \citep{Cochran_1988}. The most successful of these early efforts
\citep{Campbell_1979} introduced the concept of observing stars through a
hydrogen fluoride absorption cell to imposed a stable wavelength reference
directly on the beam of starlight prior to entering the spectrometer. This is an
improvement on the telluric O$_2$ technique as the absorption cell is much more
stable than the earth's atmosphere. Telluric O$_2$ lines vary in depth with the
altitude of the observation, and they are subject to winds, leading to a RV
uncertainty exceeding 10 \ms \citep{Cochran_1988,Smith_1982}.

Two techniques emerged in the 1990s that have resulted in nearly all the known
exoplanets discovered by precision Doppler  RV surveys. The
super-stabilized spectrometer ELODIE led to the discovery 51 Peg
\citep{Mayor_1995}. Super-stabilized spectrometers are temperature and pressure
stabilized. Pressure stabilization is achieved by mounting the spectrometer in a
vacuum chamber. Starlight is fed into the spectrometer via a fiber scrambler. A
separate fiber simultaneously feeds a wavelength reference (e.g. ThAr hollow
cathode lamps, Fabry-P\'{e}rot (FP) interferometers, laser frequency comb
(LFC)).

The iodine absorption cell technique \citep{Butler_1987,MarcyButler_1992}
resulted in five of the first six known planets
\citep{MarcyButler_1996,Butler_1996b,Butler_1997} and 170 of the first 200
planets \citep{Butler_2006}. The iodine technique does not require a
super-stabilized spectrometer with a vacuum chamber and fiber scramblers because
the reference spectrum is carried on the beam of starlight prior to entry into
the spectrometer. As iodine cell spectrometers are typically not stabilized,
they are subject to a variable wavelength scale and instrumental
point-spread-function (PSF). Since the iodine reference is embedded on the
stellar spectrum, the wavelength scale and PSF can be recovered directly from
the observation, at the expense of a more complex data reduction package
\citep{Butler_1996}. 

The iodine absorption lines are located in the wavelength range 5000~\AA\
to 6500~\AA ; outside this wavelength range they need to be used with additional
wavelength references (e.g. Th/Ar HCLs). For G and early K dwarfs, nearly all
the Doppler RV information resides in the wavelength range from 4000 to 6000~\AA\ \citep{Merline_1985}. Including the stellar  RV information in the
region below 5000~\AA\ primarily benefits the late F and early G stars. For the
late G and K dwarfs, this region adds relatively little additional  RV
information as the stellar flux is rapidly decreasing. Pushing further into the
red, beyond 6000~\AA, primarily benefits the late K and M dwarfs where the flux is
increasing. The number of telluric lines significantly increases beyond 6000~\AA,
which complicates the data reduction and decreases the information content.

The super-stabilized and iodine absorption cell techniques have subsequently
been adopted by nearly all precision Doppler spectrometers. Examples include
CORALIE \citep{Queloz_2001}, the HIRES \citep{Vogt_1994} spectrograph on the
Keck telescope, the UCLES \citep{Diego_1990} spectrograph on the Anglo
Australian telescope, UVES \citep{Dekker_2000} on the European Southern
Observatory's (ESO) VLT, HARPS \citep{Mayor_2003}, the PFS \citep{Crane_2010}
spectrograph on the Magellan telescope, and the ESPRESSO spectrograph
\citep{Pepe_2021} on ESO's VLT. Iodine cells can easily be added to existing
echelle spectrometers simply by mounting them in front of the spectrometer
entrance slit. Super-stabilized spectrometers must be custom built, mounted
inside of a vacuum chamber, temperature controlled at the 0.001 \degC\ level,
and fed by dual fiber scramblers, all of which drives up the cost and
complexity. Using an iodine cell to calibrate both the stellar and reference
fibers of a super-stabilized spectrometer potentially offers a mechanism to make
full use of the wavelength range of the spectrometer. The iodine calibration
allows for the recovery of the two major problems that currently limit the long
term stability of super-stabilized spectrometers: changes in the PSF and
relative motion of the stellar and reference fibers.

For spectrometers with a resolving power above $\approx$120K both
super-stabilized and iodine cell spectrometers achieve an uncertainty
less than 1 \ms. Newer generations of very high-precision
super-stabilized spectrographs are aiming to achieve at least an order of
magnitude lower uncertainty than this. The science goals of the ESPRESSO
spectrograph range from the detection of rocky exoplanets around solar-type
stars to placing tighter bounds on the possible variation of the fundamental
constants, and require a RV uncertainty of around 0.1 \ms. One science goal
for the ANDES \citep{ANDES} high resolution spectrograph planned for ESO's
Extremely Large Telescope (ELT) is to directly measure the acceleration of the
expansion of the Universe using the Sandage test \citep{Sandage_1962}, and this
requires a wavelength stability of 0.02 \ms over a period of 10 years. 

To avoid losing their RV zero-point, super-stabilized spectrometers
must maintain their optical, thermal and mechanical stability to extraordinarily
high tolerances. If the long term goals include finding earth-analogs or
carrying out the Sandage test, the stability must be maintained at the level of
atomic length scales for decades. This Herculean task has not been demonstrated
to date. Significant jumps in the HARPS nightly zero-point have been reported on
roughly 30 nights \citep{Trifonov_2020}. These jumps are attributed to thermal
variations and differential motion of the stellar and reference fibers (G.
LoCurto, 2025 private communication). The Keck Planet Finder, an ESPRESSO-like
super-stabilized spectrometer, has not been able to maintain its nightly
zero-point since May 2025 due to problems with the thermal control system
\citep{keck_status,keck_stability,kpf_stab_announce}. Anecdotal evidence from
ESPRESSO instrument scientists suggests that small earthquakes and other
mechanical ``jolts'' can cause differential motion of the stellar and reference
fibers resulting in RV zero-point jumps at the level of 2 \ms.
The spectrometer PSF will change whenever the optics or detector is upgraded, or
potentially whenever the vacuum chamber is opened. An example of this is the
jump in RV of $\approx$8 \ms after the HARPS detector was upgraded in 2015
\citep{Trifonov_2020}.

In October 2020, at a meeting of the ESO Working Group on Line Calibrations for
the ELT, we proposed that ESO consider incorporating an iodine cell into the
design of the ANDES spectrograph. Subsequently, it was agreed that an iodine
cell should be installed on the ESPRESSO spectrograph to provide an independent
wavelength reference that is embedded in the stellar spectrum and is thus
complementary to the LFC and FP wavelength references usually used with that
instrument. This paper describes the design and construction of the cells and
the results of the first commissioning run on ESPRESSO in May 2023.

\section{Design, Construction and Calibration of the ESPRESSO Iodine Cell} \label{sec:design}

The method of construction of suitable iodine cells for use as wavelength
references for an astronomical spectrograph is described in
\citet{MarcyButler_1992,Butler_1987}. Key to the technique is that the cells
must be sized correctly for the spectrograph and that quantity of iodine in the
cell should be such that all of it can be vaporized by heating the cell up to
roughly 50~$^\circ$C. By restricting the quantity of iodine in this way, the
vapor density of iodine in the cell, and thus the spectrum, is insensitive to
the temperature. This precludes the use of typical commercial iodine cells used
for laser spectroscopy, which are often over-filled and used at ambient
temperature \citep{Crause_18}.

The ESPRESSO spectrograph can take light from any of the four unit telescopes of
the VLT, with the light from the telescopes directed through the Coud{\'e}
trains to the atmospheric dispersion compensator (ADC) and front ends in the
Combined Coud{\'e} Laboratory (CCL)\citep{Pepe_2021}. The front ends contain
beam condition and mode selection optics to direct the light into optical
fibers, and the optics to insert the calibration light from the LFC, Th/Ar
lamps, and the Fabry-P{\'e}rot units. Sufficient room exists before the ADC to
insert an iodine cell and this is where the cells were positioned. Although
light can be fed into the ESPRESSO spectrograph from any of the unit telescopes,
for the purpose of this experiment, the iodine cell was set up before the ADC
for UT2 for all of the observations where it was used. At this position, the
light from the telescope is a F/22.8 converging beam and has a diameter of
roughly 50~mm at a position about 100~mm before the entrance of the ADC.

% -- Where to put the cell, and how that set the dimensions.

Four iodine cells made of Pyrex were fabricated at Allen Scientific Glass
and coated at Evaporated Coatings Inc.\footnote[1]{Certain commercial equipment
and materials are identified in this article to adequately specify the
experimental procedure. Such identification does not imply endorsement by the
National Institute of Standards and Technology, nor does it imply that they are
the best available for the purpose}. The diameter of the cells was 75~mm, giving
a clear aperture of about 50~mm. The length was 100~mm and each window was
6.5~mm thick. The four cells were filled at temperatures of 35~$^\circ$C
(AS-31), 36~$^\circ$C (AS-34), 37~$^\circ$C (AS-33), and 38~$^\circ$C (AS-32).
The cells were coated with a hard, broad-band anti-reflection coating with
a reflectivity of $<$ 0.5~\% from 5000~\AA\ to 6500~\AA. The coating curve from
the manufacturer covers the region 4000~\AA\ to 7400~\AA\ and shows a rapid rise
in the reflectivity at shorter wavelengths, from 0.5~\% at 4400~\AA\ to 5~\% at
about 4050~\AA.

\begin{figure}
\plotone{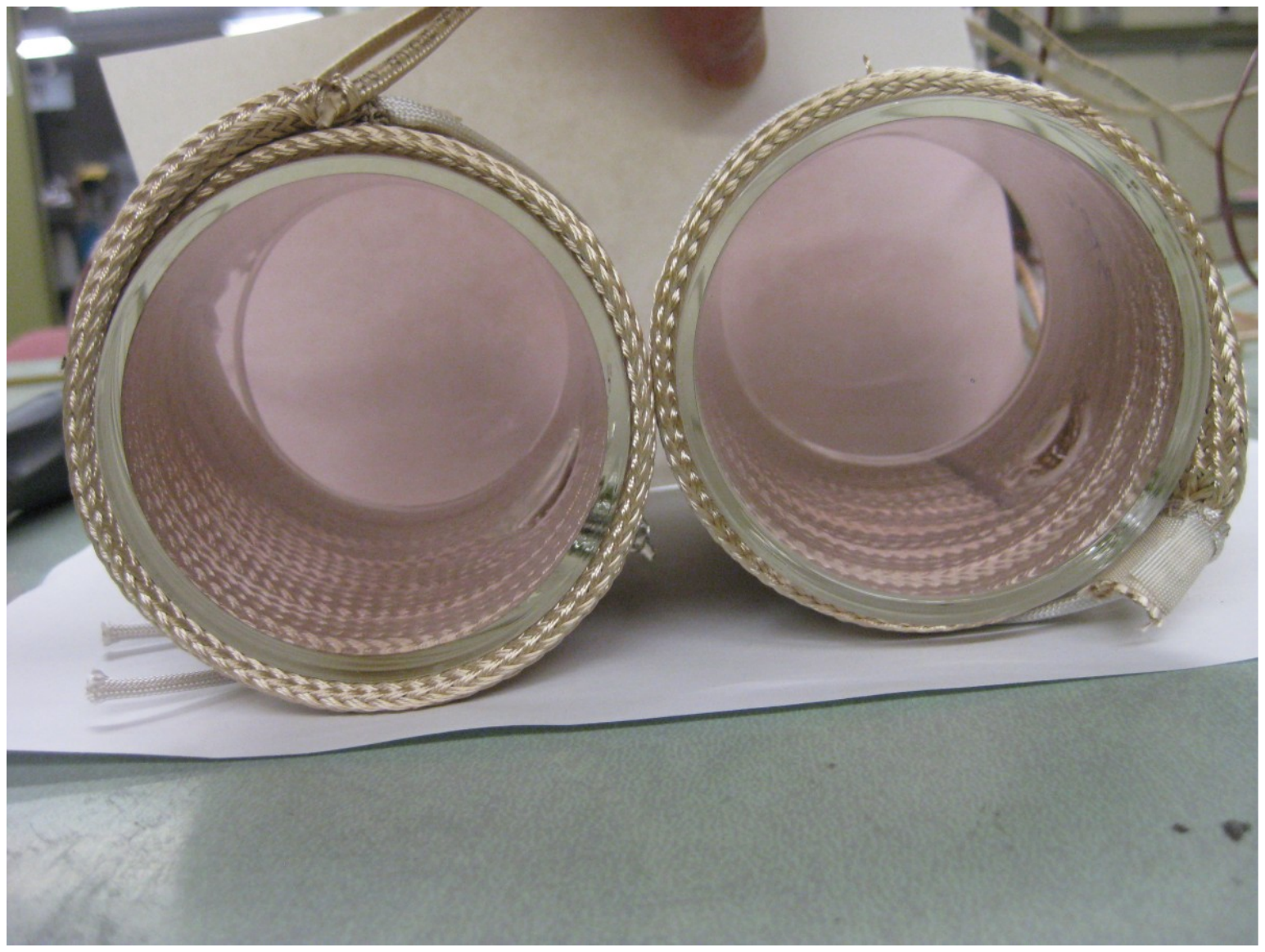}
\caption{Photo of iodine cells \label{Fig:cell_photo}}
\end{figure}

A photo of two of the cells is
shown in Figure \ref{Fig:cell_photo}, taken with each cell heated to
50~$^\circ$C. For operation with a spectrograph, each cell is heated to
60~$^\circ$C for at least half an hour before the start of the measurements to
ensure that all the iodine was fully vaporized. After this, the cells are
checked visually to ensure that no specks of solid iodine remained. An iodine
cell filled with the correct amount of iodine will have a light pink color at
the operating temperature. All four of the cells have a light pink color with an
optimal fill pressures of iodine. All four cells would be suitable for
calibration of an astronomical spectrograph. One empty cell (labeled AS-35) of
the same dimensions was also fabricated that contained no iodine. This was
constructed to compensate for any shift in the focus of the UT2 telescope due to
the windows of the iodine cell. However, during the commissioning run, it was
found that the focus shift was very small and this cell was thus not used.

\subsection{Iodine atlases}\label{sec:atlases}

The transmission of the cells was measured with the NIST 2-m Fourier transform
spectrometer (FTS). The background source was a 1000~W Xe lamp that emits a
strong continuum from below 2000~\AA\ to over 2.5~$\mu$m. The iodine spectrum is
concentrated in a region from about 5000\AA\ to 6200~\AA, and it is important
that the spectrum illuminating the cell is restricted to that region in order to
obtain a good signal-to-noise ratio (S/N). Three filters were thus used: a hot
mirror, transmitting from 4000~\AA\ to 7000~\AA; a GG495 filter; and a BF40
filter. The lamp was roughly focused on the entrance aperture of the FTS, but
this focusing is not critical. The background spectrum of the xenon lamp and
three filters is shown in Figure \ref{Fig:Xe_lamp}, together with the
transmission of the empty cell (see below).  

\begin{figure}
\plottwo{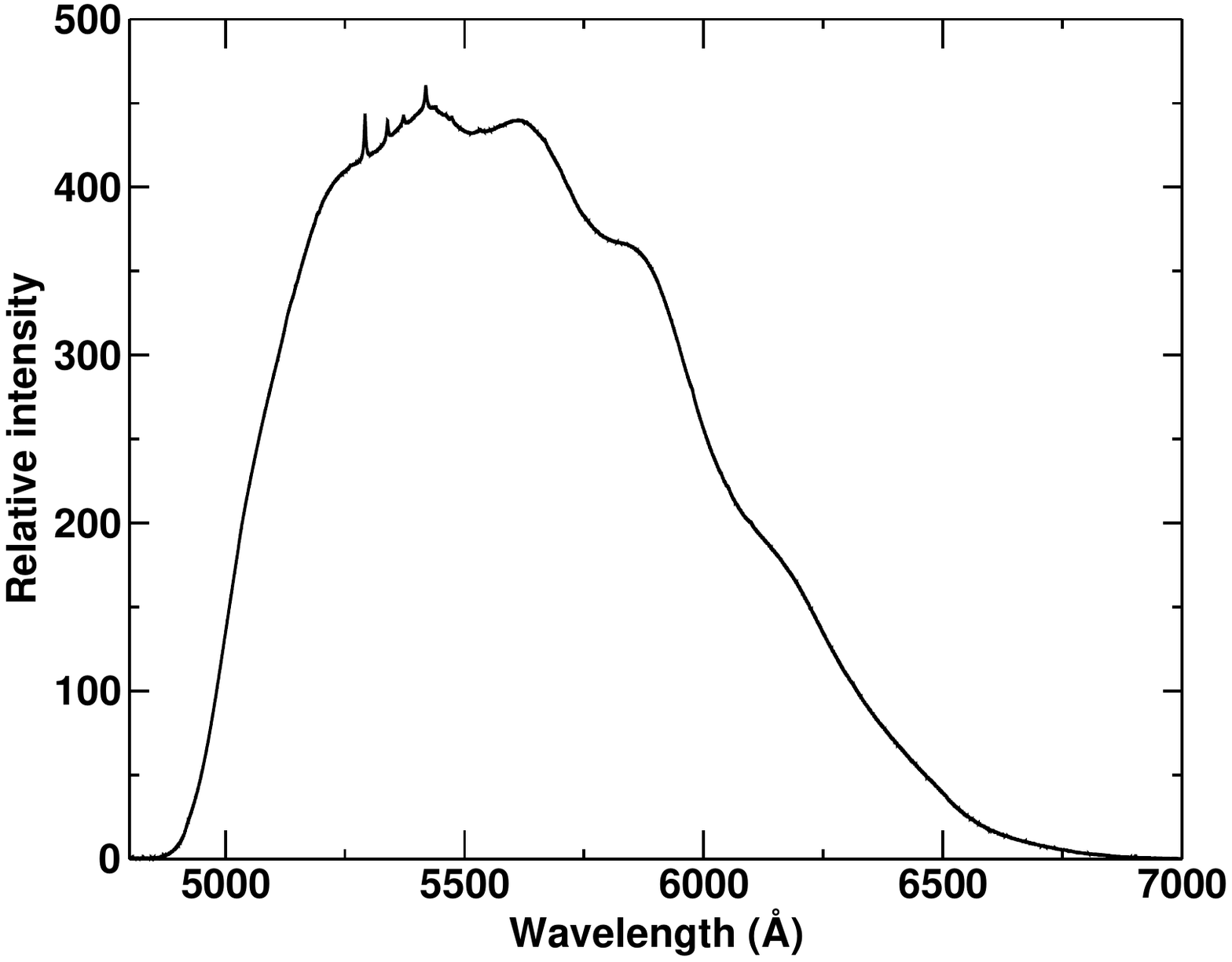}{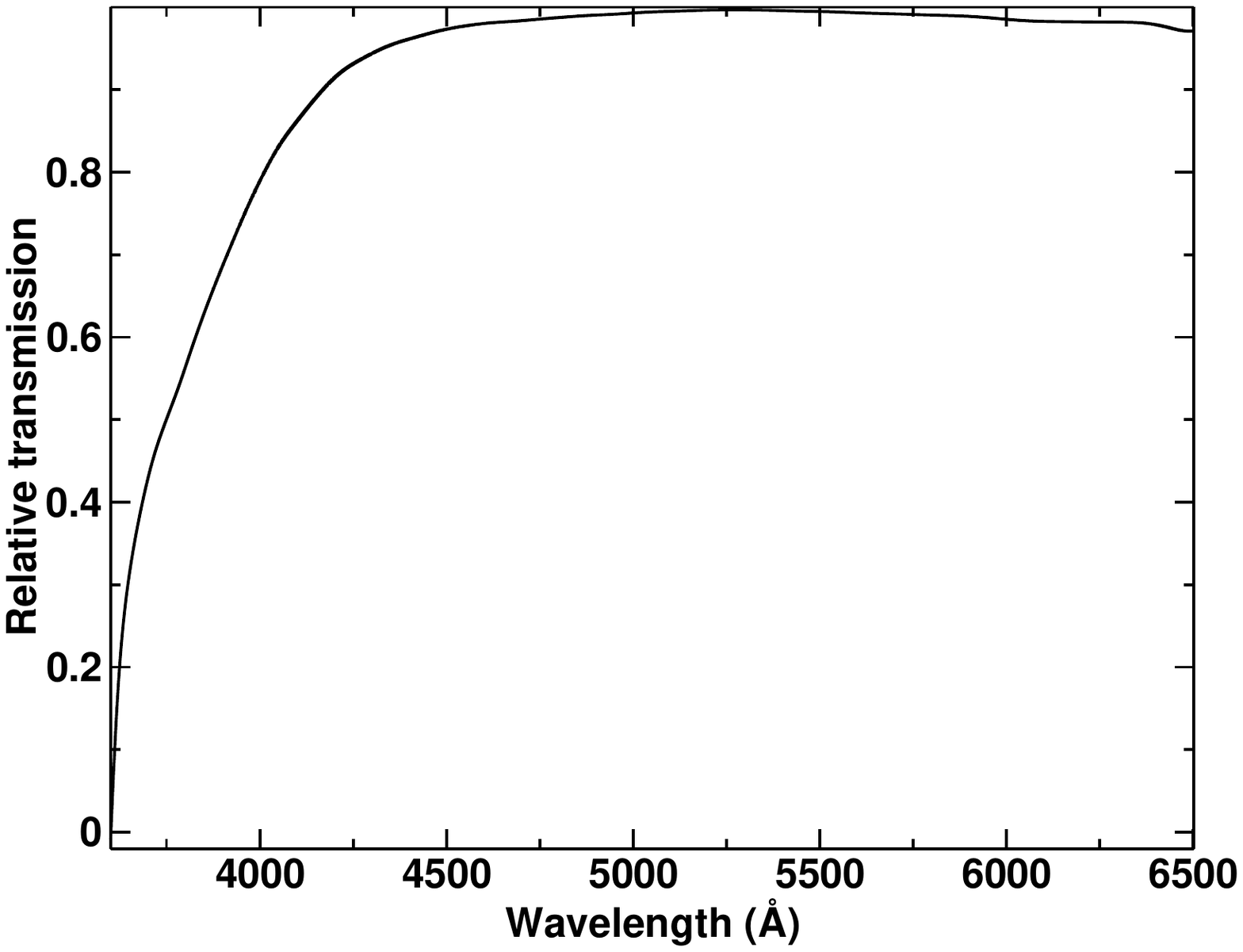}
\caption{Left: Spectrum of the xenon lamp and filters used to illuminate the 
cells. Right: transmission of the empty cell derived from spectra 18 and 19 in 
Table \ref{Tab:FTS_meas}. The rapid drop in the transmission below 4000~\AA\ corresponds to the 
rise in reflectivity of the coating curve supplied by the coating manufacturer.
\label{Fig:Xe_lamp}}

\end{figure}

Each cell was measured at two different resolutions of 0.018~\cm\ and 0.01~\cm\,
corresponding to resolving powers of roughly 1 million and 1.8 million at the
center of the absorption band. The wavenumber region for all measurements with
the 2-m FTS was 14200~\cm\ to 21300~\cm\ (about 4700 – 7040~\AA ). The AS-33
cell was also measured at a resolution of 0.007~\cm, corresponding to a
resolving power of about 2.5 million. Each spectrum took about an hour to
measure and gave a S/N of between 600 and 1700, depending on the resolution.
Some measurements were repeated after it was noticed that the imaginary part of
the spectrum had started to increase during the day. This suggested that there
was a drift in the alignment of the FTS, causing an asymmetry in the
interferogram on each side of zero path difference and thus an imaginary
component to the spectrum. The full set of measurements is given in Table
\ref{Tab:FTS_meas}.

The initial measurements of the empty cell were taken several hours after the
measurement of the background lamp spectrum, and it was clear that the FTS had
drifted between the two sets of measurements. A spectrum of the cell was thus
taken using the NIST vacuum-ultraviolet (VUV) FTS at a resolution of 128~\cm,
using a tungsten lamp as a background source, in order to obtain a better
measurement of the transmission of the empty cell. Since the spectrum was taken
at a larger resolution, it was not necessary to use optical filters to obtain an
adequate S/N. The wavenumber region used for these measurements was 0 to
28000~\cm. The transmission of the empty cell is shown in the right panel of
Figure \ref{Fig:Xe_lamp}.  The rapid drop in the transmission of the cells
below 4000~\AA\ corresponds to the rise in the reflectivity of the coating
curves that were provided by Evaporated Coatings Inc., as noted above.

\begin{table}
%\centering
\caption{Summary of the iodine cell measurements with the NIST 2-m FTS}
\label{Tab:FTS_meas}
\begin{tabular}{llllllll}
\tableline
No. & Cell   & Fill           & File Name  & No.   & Resolution & Peak & Imaginary \\
 &        & Temperature    &            & scans & ~\cm\ & S/N  & part \\
   \tableline                                                
1  & AS-35  & empty          & Xe032023.002 & 4   & 0.018  & 400 \\
2  & AS-35  & empty          & Xe032023.003 & 16  & 0.018  & 760 \\
3  & AS-31  & 35~\degC\ & I2032023.001 & 32  & 0.018  & 1500 & small \\
4  & AS-32  & 38~\degC\ & I2032023.002 & 32  & 0.018  & 1500 & increased  \\
5  & AS-33  & 37~\degC\ & I2032023.003 & 27  & 0.018  & 1300 & large  \\
6  & AS-34  & 36~\degC\ & I2032023.004 & 32  & 0.018  & 1300 & large  \\
   \\                                                    
7  & AS-34  & 36~\degC\ & I2032123.001 & 32  & 0.018  & 1400 & increased  \\
8  & AS-34  & 36~\degC\ & I2032123.002 & 11  & 0.018  & 1000 & reduced \\
9  & AS-34  & 36~\degC\ & I2032123.003 & 32  & 0.018  & 1700 & small  \\
10  & AS-33  & 37~\degC\ & I2032123.004 & 9   & 0.018  &  370 & small  \\
11  & AS-33  & 37~\degC\ & I2032123.005 & 32  & 0.018  & 1150 & small  \\
12  & AS-33  & 37~\degC\ & I2032123.006 & 16  & 0.010  &  600 & small  \\
   \\                                                     
13  & AS-33  & 37~\degC\ & I2032223.001 &  6  & 0.007  &  400 & increased  \\    
14  & AS-33  & 37~\degC\ & I2032223.002 & 21  & 0.007  &  600 & reduced  \\   
15  & AS-34  & 36~\degC\ & I2032223.003 & 18  & 0.01   &  950 & small  \\
16  & AS-31  & 35~\degC\ & I2032223.004 & 10  & 0.01   &  700 & small  \\
17  & AS-32  & 38~\degC\ & I2032223.006 & 22  & 0.01   & 1000 & small  \\
 \\                                                        
18  & none\tablenotemark{a}  &       & W041923.001 & 128 & 1.0 & 750  &  \\
19  & AS-35\tablenotemark{b} & empty & W041923.002 & 128 & 1.0 & 720  &  \\
\tableline    
\end{tabular}
\tablenotetext{a}{Background spectrum of a tungsten lamp taken with NIST VUV 
FTS.}
\tablenotetext{b}{Spectrum of tungsten lamp taken through the empty cell with 
NIST VUV FTS.}
\end{table}

% \subsection{FTS spectra of iodine cells \label{sec:FTS_spectra}}

\begin{figure}
\plotone{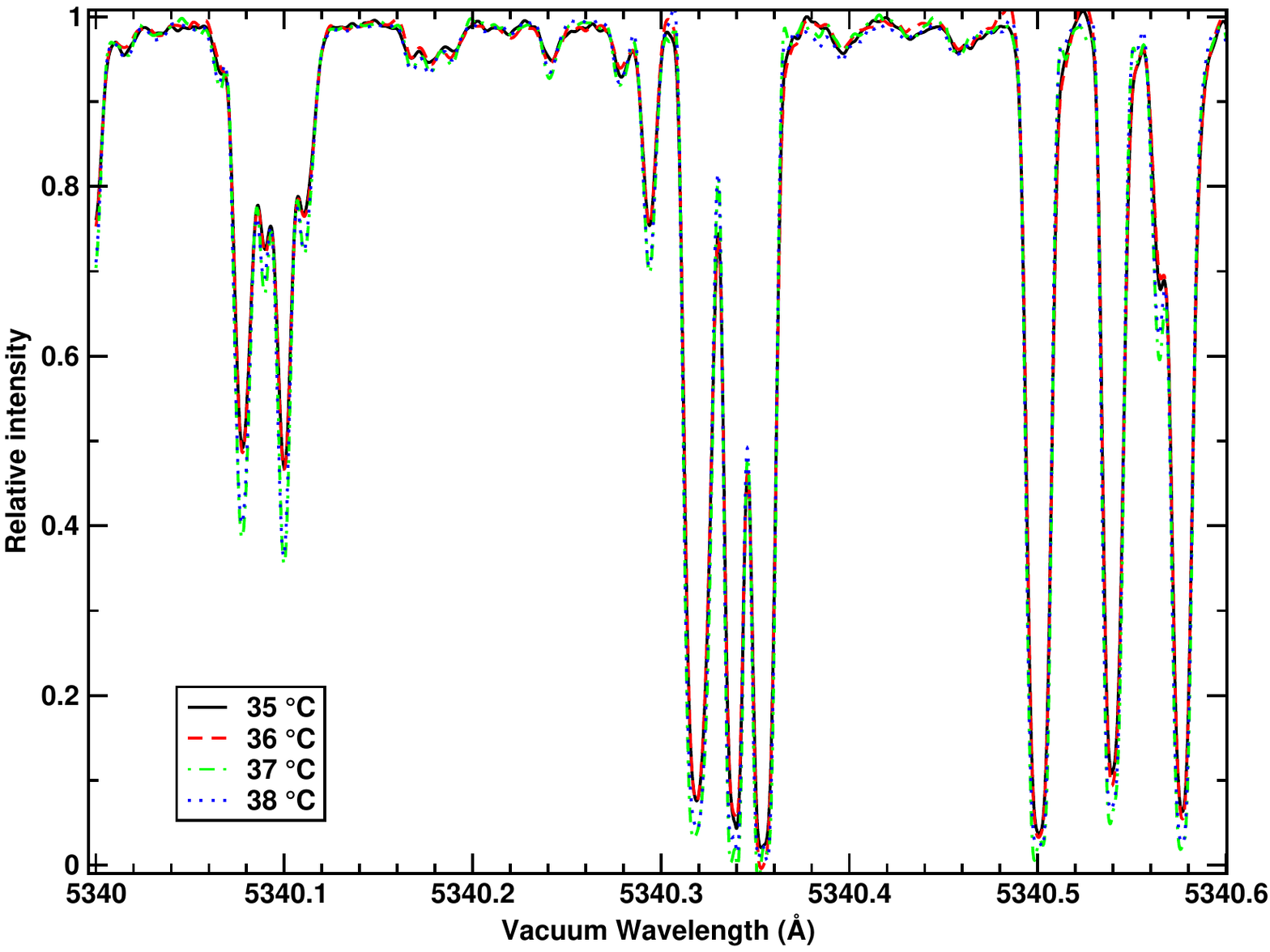}
\caption{Comparison of the scans for each cell at a resolution of 0.01~~\cm. 
Turquoise: AS-34; Blue:AS-31; Red:AS-32; Green:AS-33. 
\label{Fig:cell_comparison}}
\end{figure}

A comparison of the scans of the iodine cells is shown in Figure
\ref{Fig:cell_comparison}. All four cells have very similar spectra, indicating
that the actual fill pressures inside the cells did not vary much. All four
would be excellent wavelength references for use on an astronomical
spectrograph.

\begin{figure}
\plottwo{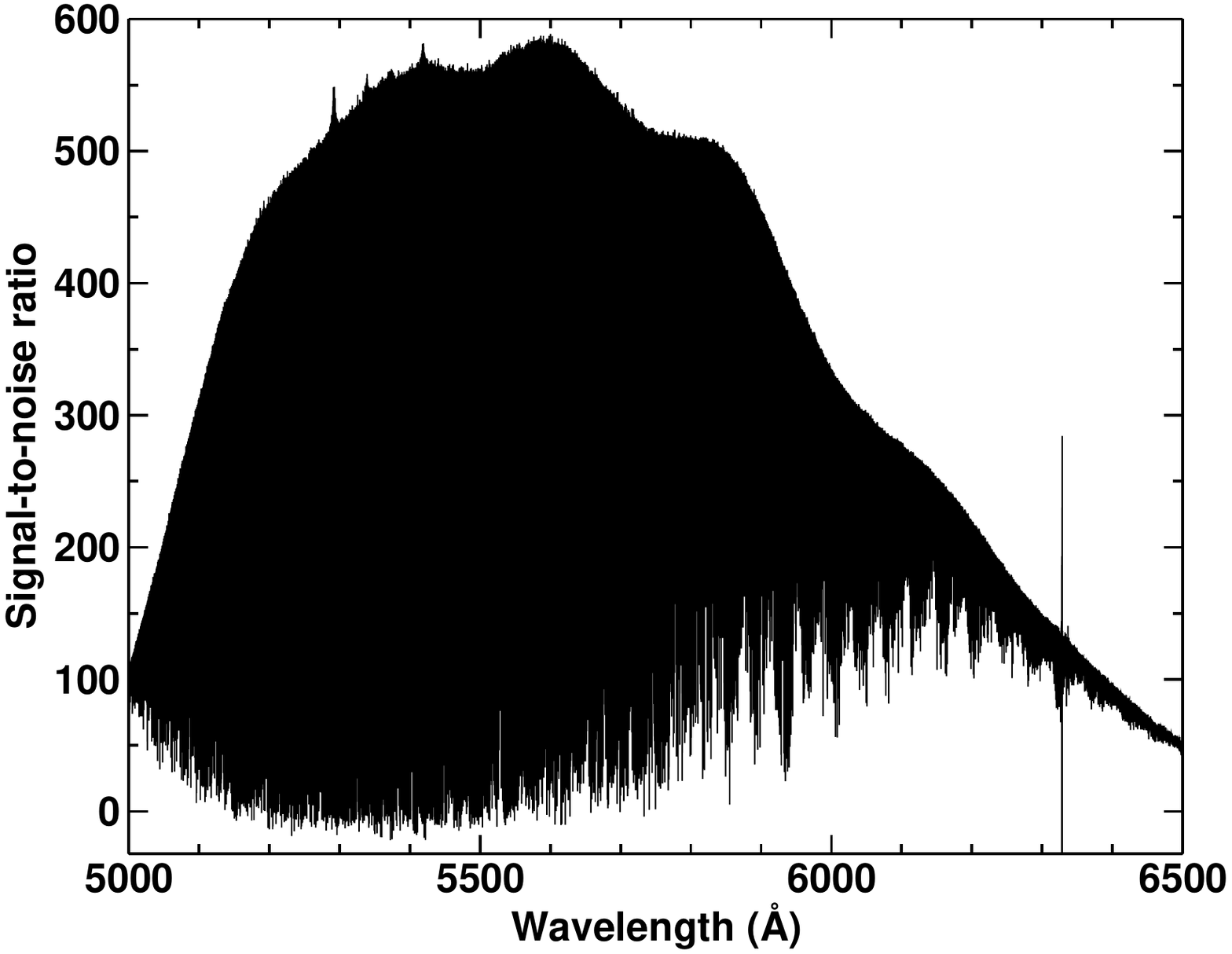}{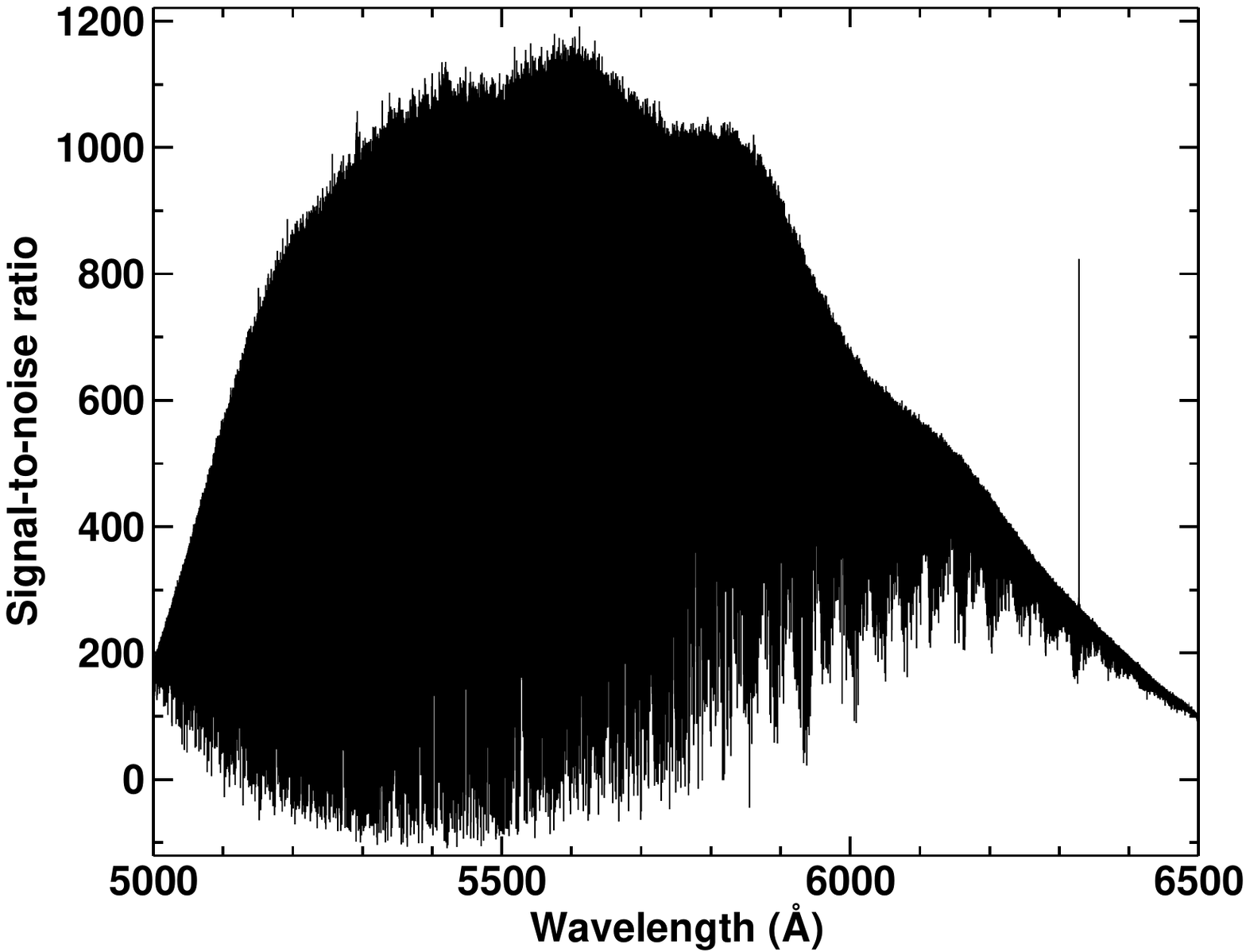}           
\caption{Comparison of FTS scan of cell AS-33 at a resolution of 0.007 \cm\ 
(left) and 0.018 \cm\ (right). \label{Fig:AS-33}}
\end{figure}

Figure \ref{Fig:AS-33} shows the spectrum of cell AS-33 scanned at a resolution
of 0.007~\cm\ and at a resolution of 0.018~\cm. Although the measured S/N is
higher at a resolution of 0.018~\cm, the visual appearance of the spectrum
appears to be noisier. This is due to the instrument function of the FTS, which
is a sinc function. This function is convolved with every iodine line in the
spectrum, resulting in lines that appear to go below zero, and lines that appear
to poke up above the background continuum. When the FTS template spectrum is in
turn convolved with the instrument function of ESPRESSO, these sinc functions
vanish, giving a much smoother appearance.

Based on these results, we selected the cells AS-33, filled at 37~$^\circ$C, and
AS-34, filled at 36~$^\circ$C, together with AS-35, the empty cell, to send down
to Paranal for the ESPRESSO measurements. These cells have been scanned multiple
times on the NIST 2-m FTS at different resolutions and with varying levels of
the imaginary component of the spectrum. Cell AS-33 was used for all the
measurements with ESPRESSO reported in the current paper.

\section{Installation of cells on ESPRESSO}

\begin{figure}
\plotone{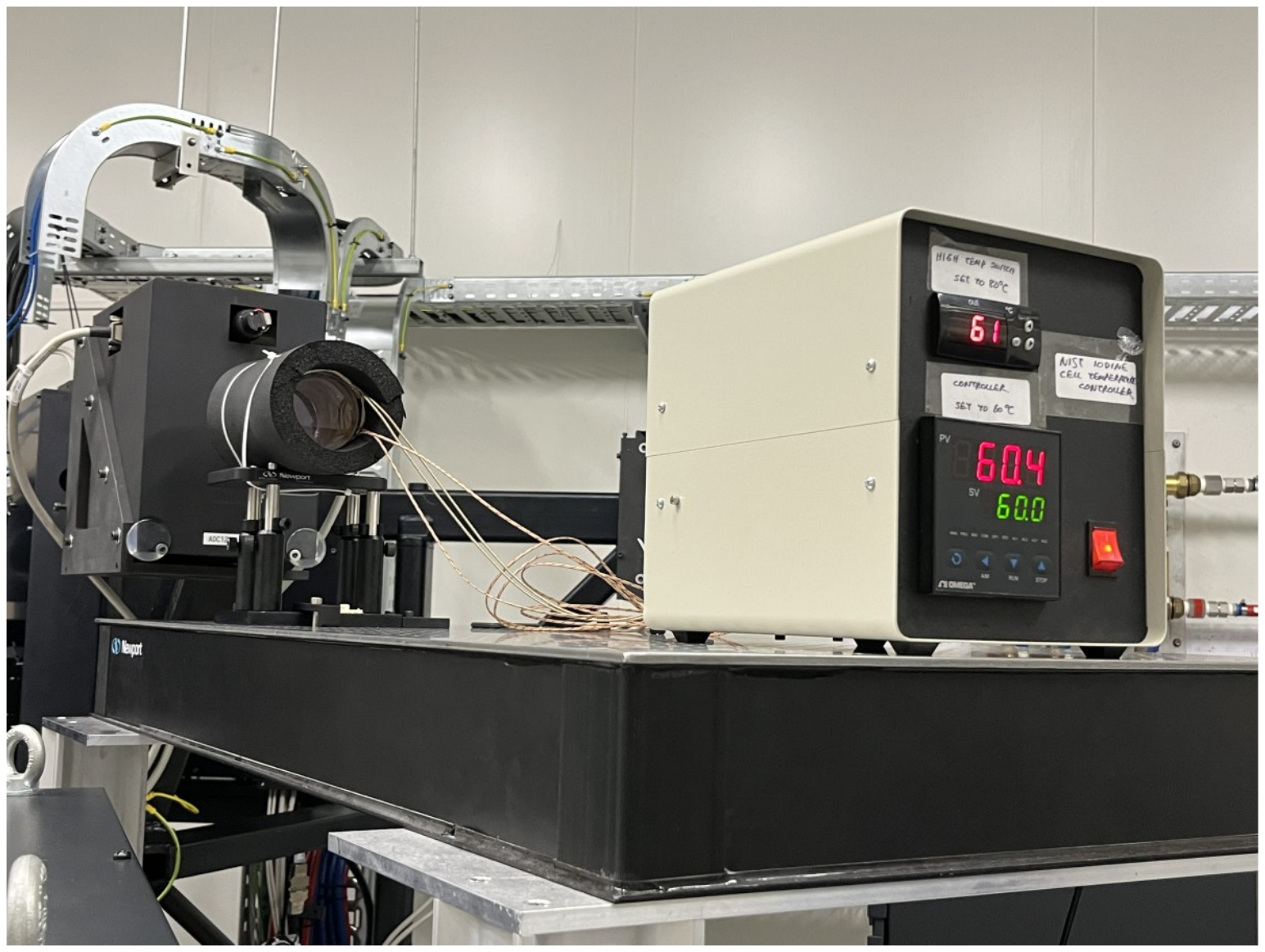}
\caption{Iodine cell AS-33 installed in front of the ADC for UT-2 \label{I2_ESPRESSO_photo}}
\end{figure}

For the initial installation of the cells on ESPRESSO, it was decided to move
the cells manually and to use a simple temperature controller to heat the cells.
Each cell was wrapped with 240~V fiberglass insulated heat tape, and two
thermocouples were taped to the cell near each window. The cells were then
covered in two layers of aluminum foil, followed by 19~mm thick flexible rubber
foam pipe insulation. The heat tape and one of the thermocouples was connected
to a standard temperature controller. Iodine cells have a very stable spectrum
over time, but can be damaged by heating to over 100~$^{\circ}$C for several
hours, which can drive the iodine vapor into the walls of the cell, reducing the
iodine density and thus changing the spectrum. This can happen with a failure of
the temperature controller. Hence a second temperature controller was connected
to the other thermocouple, and wired to turn off the power to the first
controller if the temperature of the cell exceeded 80~$^{\circ}$C. A simple
mount for the cell was constructed from standard optical hardware and clamped to
an optical table, with aluminum table clamps screwed to the table to enable the
cell to be removed and replaced reproducibly. A photo of the setup is shown in
Figure \ref{I2_ESPRESSO_photo}.

A few minor problems were found when the cell was installed on ESPRESSO. An
initial failure of the temperature controller resulted in short-term
over-heating of the cell. This was traced to a loose wire and had no long-term
consequences. The CCL room was slightly cooler than the laboratory at NIST where
the cells had been measured, and a speck of iodine was noticed on one of the
windows a day after the initial setup. This was solved by re-wrapping the cell
and placing aluminum foil under the insulation. The insulation also extended a
few cm in front of each window to minimize heat loss from the center of the
window. After placing the aluminum foil over the cells, the two temperature
controllers gave inconsistent temperatures. This was traced to a piece of
aluminum foil covering one of the thermocouples and was solved by pulling the
foil away from the thermocouple and covering it in Kapton$^1$ tape. Once these
problems had been solved, the cell worked flawlessly for the rest of the run.

\section{Observations of iodine cells with ESPRESSO}

The spectrum $I_{obs}$ of a star observed through an iodine cell with an
astronomical spectrograph as a function of the wavelength, $\lambda$, is given
by equation 1 of \cite{Butler_1996}:
\begin{equation}\label{eqn:I2}
I_{obs}(\lambda) = k[T_{I2}(\lambda)I_s(\lambda+\Delta\lambda)]\star \mbox{PSF}
\end{equation}
where $k$ is a normalization factor and $\Delta\lambda$ is the Doppler shift.
The iodine transmission function, $T_{I2}(\lambda)$, is the function measured
using the NIST FTS, as described in section \ref{sec:atlases}. The high
resolving power and S/N of this spectrum means that the contribution of the FTS
measurements to the total uncertainty is very small and can be essentially
ignored.

Ideally the intrinsic stellar spectrum, $I_s(\lambda)$, would also be registered
with a FTS at a resolution of a million-plus and a S/N $>$ 1K. Unfortunately
high resolution FTS spectra require extremely bright sources. The entire catalog
of high resolution FTS spectra of astronomical sources include the Sun
\citep{Kurucz_1984} and a handful of the brightest giant stars. For the
observations with ESPRESSO, we were able to use the Ultra High Resolution (UHR)
mode to obtain template spectra, $I_s$, at a resolution of 194,000. At the
exceptional resolving power of the UHR mode, the deconvolution contributes only
a small fraction to the total error budget.

The PSF in equation \ref{eqn:I2} is measured by observing the spectrum of a
bright, rapidly rotating OB star through the iodine cell. The intrinsic spectrum
of these stars is essentially featureless, and thus act as a continuum source to
observe the iodine cell spectrum with ESPRESSO. The observed spectrum is thus
the convolution of the iodine spectrum, obtained from the FTS spectrum, with the
ESPRESSO PSF and thus the PSF can be obtained by deconvolution, as described in
section 2.2 of \cite{Butler_1996}. One complication is that the observed
spectrum of the OB star also contains telluric lines; these are mapped by
observing the OB star without the iodine cell and these pixels are avoided in
the subsequent analysis.

\begin{figure}
\centering
\includegraphics[scale=.5,angle=180]{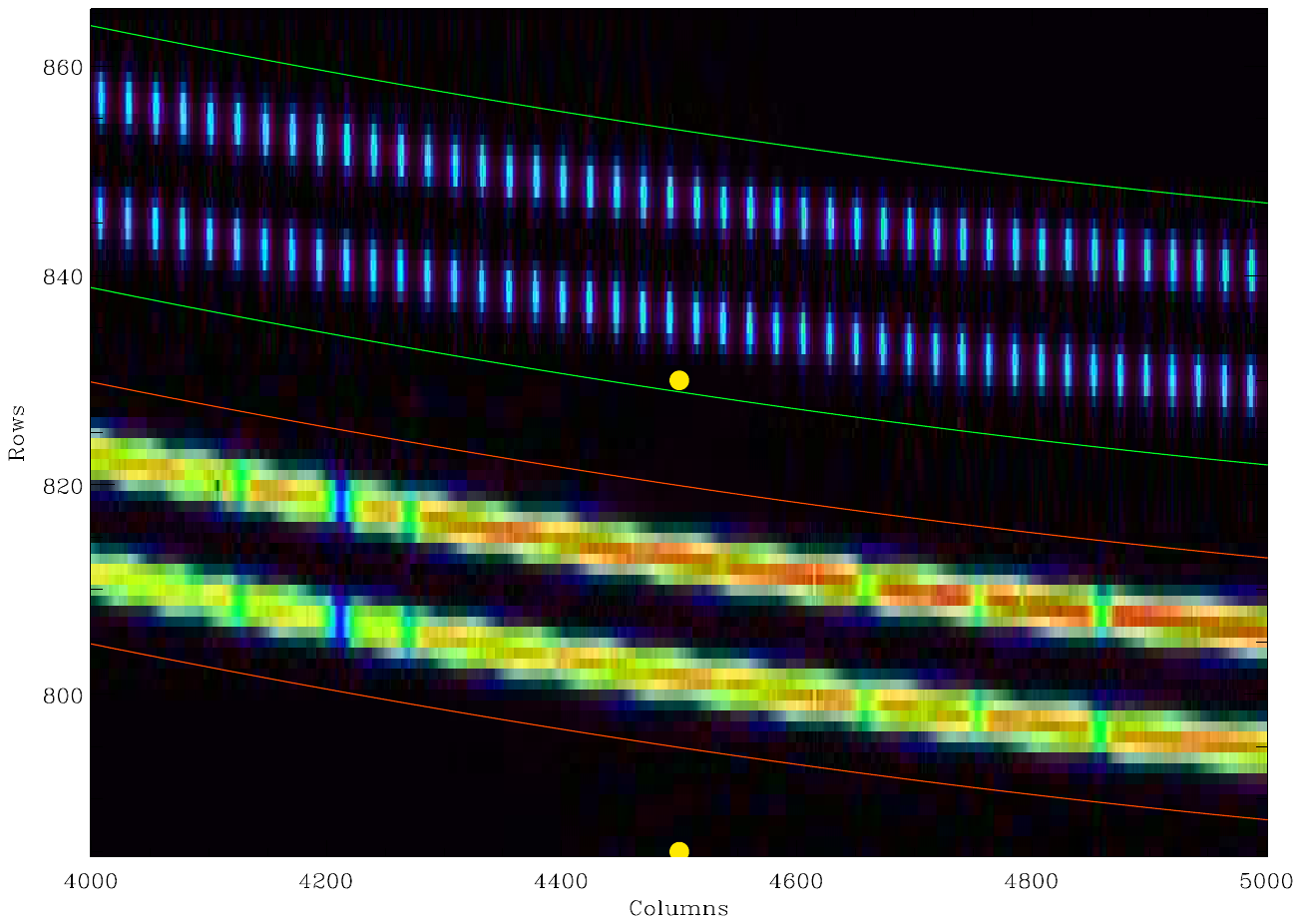}
\caption{Bias subtracted raw image of HD102365 in the middle of the iodine
region. The stellar (lower) and Fabry-P\'{e}rot reference (upper) are each split
into two slices. Each of the slices are separated by roughly 11 
pixels. The boundaries of the stellar and reference regions are shown as red and 
green lines respectively. \label{fig:raw}}
\end{figure}

The raw reduction of ESPRESSO data has an extra complication. The reference and
stellar source are each fed with a fiber sliced into two parts by a pupil
slicer (See Fig. 4 of \citet{Pepe_2021}). Figure \ref{fig:raw} shows a bias
subtracted 1000 pixel section of the raw image of an order in the middle of the
iodine region from an observation of the bright G5 dwarf HD102365. The reference
(top) and stellar source (bottom) are each divided into two parallel bands. Each
of the fiber slices are offset by roughly 11 pixels perpendicular to
dispersion.

\begin{figure}
\centering
\includegraphics[scale=.5,angle=180]{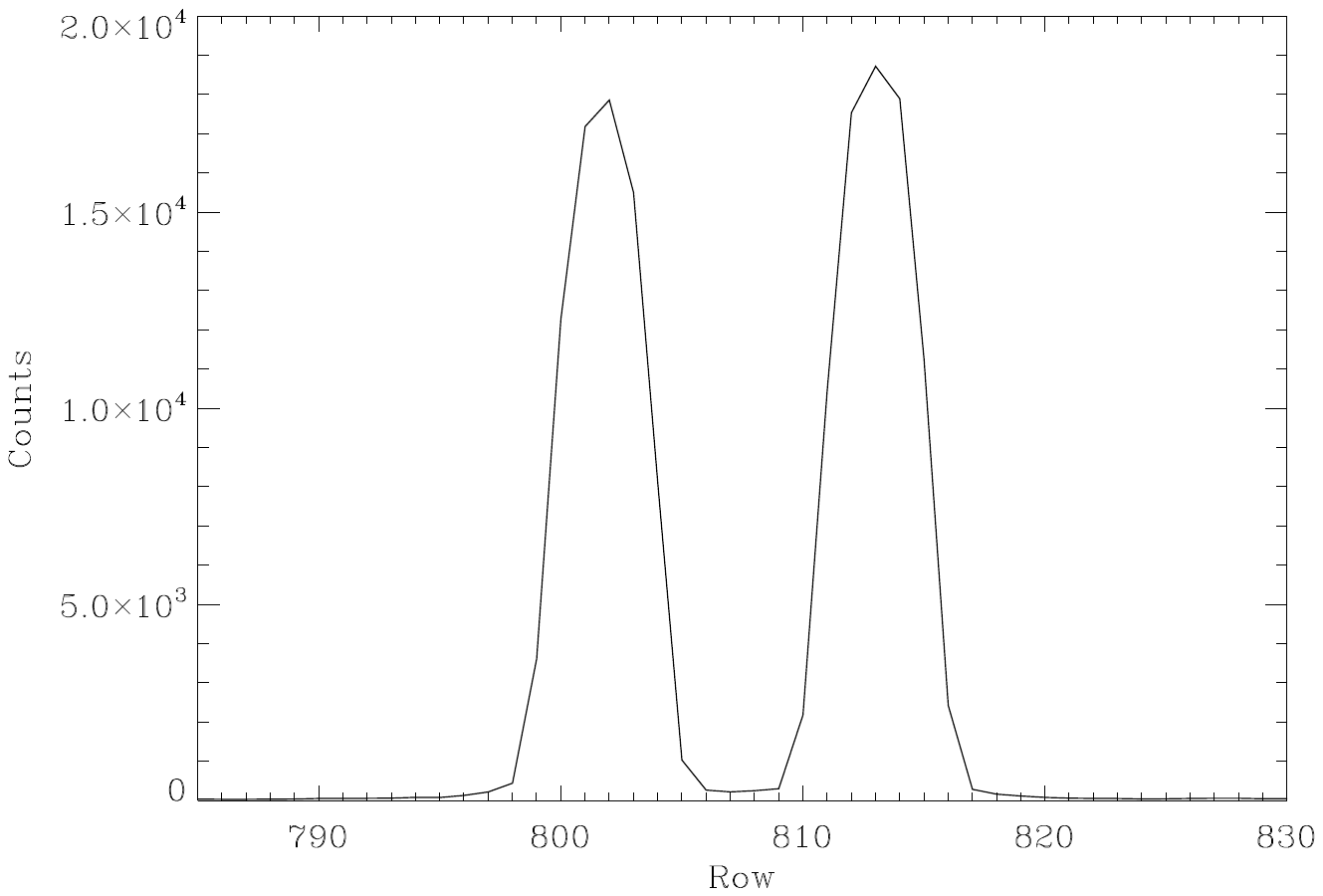}
\caption{Perpendicular cut through a stellar order of the raw image of HD102365.
The orange dots in the previous figure show the lower and upper boundaries of
the cut. The starlight is carried on two slices of the fiber separated by 11 pixels. The
perpendicular broadening function approaches zero about 10 pixels from the
maximum of each fiber. The wings of the broadening function overlap in the
region between the peaks of the two fibers. \label{fig:raw_cut}}
\end{figure}

Figure \ref{fig:raw_cut} shows a perpendicular cut through the two stellar 
slices at column 4500. The location of the cut is indicated by the two orange
dots in the previous figure. The 11 pixel perpendicular offset in the two 
slices is immediately obvious. The broadening function perpendicular to
dispersion approaches zero about ten pixels from the peak of each  slice.
The two  slices would need to be separated by about 20 pixels to prevent
any overlap on the CCD. The overlap is visible in the four pixel section (pixels
806 to 809) in the figure. This would be of little consequence if the two 
slices were perfectly aligned in the dispersion direction. Unfortunately they
are offset by rough 0.6 pixels. The necessitates extracting and analyzing each
slice separately. Each raw image therefore generates two sets of reduced
stellar and reference spectra. To minimize the ``cross-talk'' between the two
slices, we assign zero weight to the two middle pixels in the overlap
region (pixels 807 and 808) in the final extraction.

The Doppler RV analysis is carried about by dividing the reduced spectra
into 200 pixel chunks in the iodine region, corresponding to about 1.8~\AA,
allowing the PSF to vary over the format. Each chunk-set provides an independent
RV-set. There are roughly 565 useful chunks in each spectrum. As each raw
image generates two spectra, there are a total of 1130 RV sets used to
generate the final weighted mean RV for each observation. The internal
uncertainty is derived from the scatter of the individual RV sets. The
internal uncertainty does not account for systematic errors or stellar jitter.
 
\subsection{ESPRESSO PSF} \label{sec:PSF}

The single most important attribute of a precision Doppler RV spectrometer
is resolving power (R). The Doppler RV information in a stellar spectrum
lies entirely in the slopes of the stellar absorption lines. Observing at lower
resolution degrades the line slopes and diminishes the Doppler RV
information.

\begin{figure}
\centering
\includegraphics[scale=.5,angle=180]{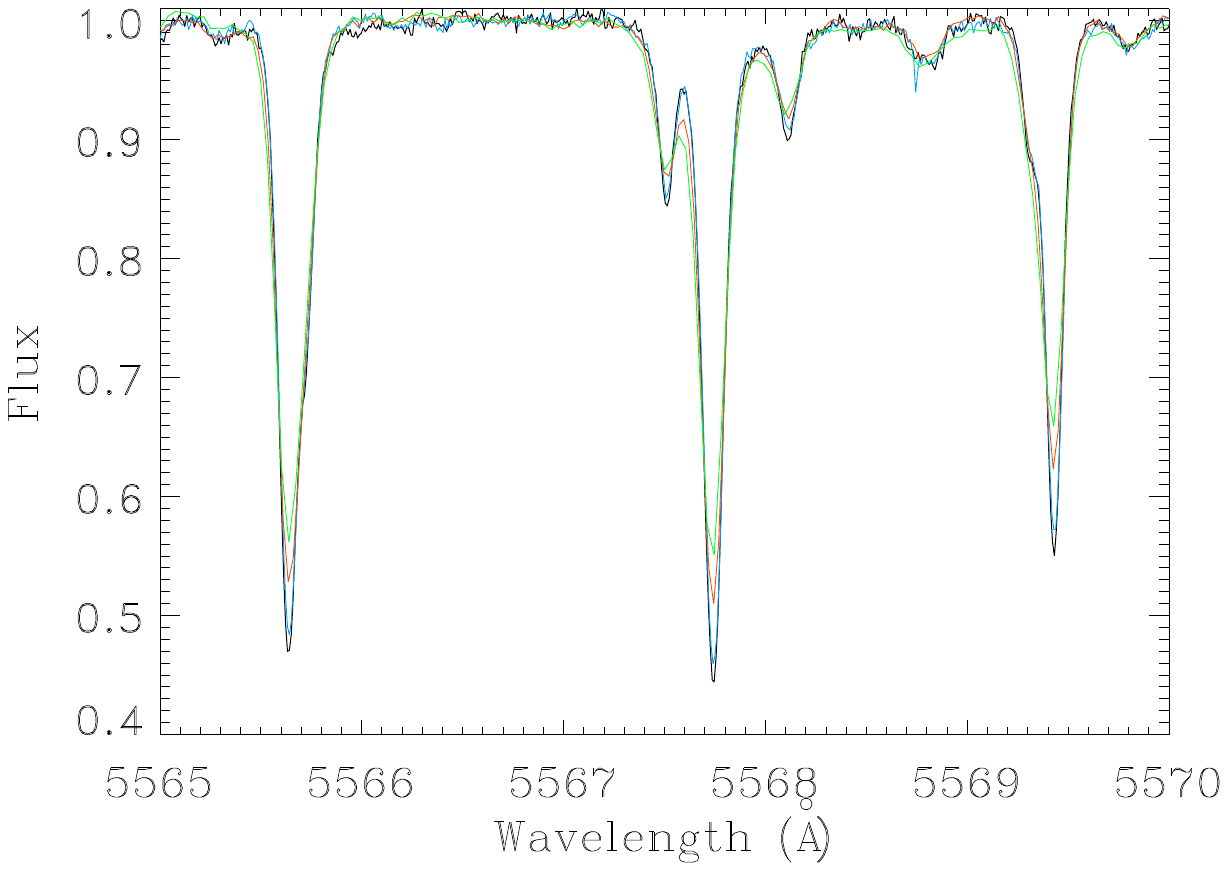}
\caption{Spectrum of HD102365, a slowly rotating G5 dwarf, at a resolving power
of 194K (black, ESPRESSO UHR), 136K (blue, ESPRESSO HR21), 82K (red, Keck 
HIRES),
and 65K (green, AAT UCLES). 
\label{fig:UHR}}
\end{figure}

The advantage of higher resolving power is shown in Figure \ref{fig:UHR}. The
spectrum of a slowly rotating G5 dwarf (HD102365) is shown at four different
resolution values. Black was taken in the ESPRESSO UHR mode with a R=194K. Blue
is the ESPRESSO HR21 (standard) mode, with R=136K. Red is Keck HIRES with a
R=82K (0.574" slit). Green is the AAT UCLES with R=65K (0.5" slit). For the
deepest line shown (5567.7~\AA ), the UHR mode produces a 3\% deeper line
compared to the ESPRESSO HR21 mode, 12\% deeper than HIRES, and 20\% deeper than
UCLES. Slowly rotating stars begin to be fully resolved at R$>$120K. Going from
a resolution of 136K to 194K results in a line depth that is only incrementally
improved. Observing at a lower resolution results in significantly shallower
lines because the broadening function of the spectrometer PSF is large compared
to the intrinsic width of the stellar line.

While stellar Doppler information is only marginally enhanced by observing
at a resolution greater than $\approx$120,000, there are at least two advantages
to observing at significantly higher resolution. As the PSF becomes narrower,
changes in the PSF have a smaller effect in generating systematic RV
offsets. The PSF of an ideal precision RV spectrometer would be a delta
function. The second advantage lies in distinguishing true Doppler RV
shifts from stellar jitter. Unlike jitter, Doppler RV shifts due to
orbiting planets do not change the shape of stellar absorption lines.
\citet{Deming_2024} have shown that the effects of the Solar magnetic cycle on
absorption line profiles can be detected and calibrated by accurately measuring
line bisectors. Accurate measurement of line bisectors can only be done at high
resolution \citet{Lohner-Bottcher_2019}.

\begin{figure}
\centering
\includegraphics[scale=.5,angle=180]{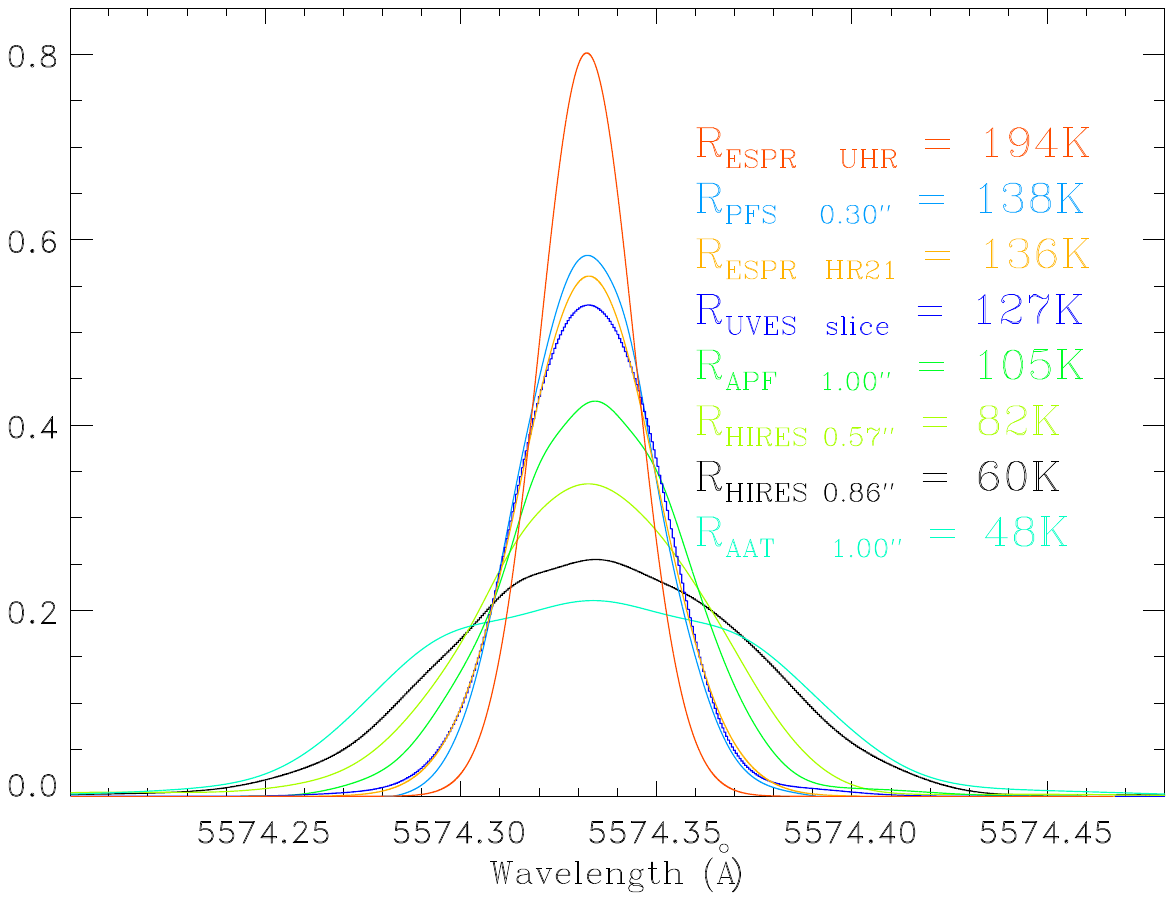}
\caption{The PSF of echelle spectrometers
with an iodine cell, in the middle of the iodine absorption region.
In each case the PSF is from an observation of a star, taken through
the iodine absorption cell.  The red PSF is from the ESPRESSO Ultra
High Resolution mode.  Orange is the ESPRESSO standard observing mode.
The slit width and resolving power, R, of each spectrometer is listed
in the figure.  The UVES observations were fed by a slicer, with an
equivalent slit width of 0.3 arc seconds. 
\label{fig:PSF}}
\end{figure}

Figure \ref{fig:PSF} shows the PSF of ESPRESSO compared with other iodine cell
equipped spectrometers. All the PSFs show small asymmetries that are more
noticeable in the lower resolution spectra. Both the Anglo-Australian UCLES
\citep{Diego_1990} and the Keck HIRES spectrometers \citep{Vogt_1994} were
designed and commissioned prior to the discovery of extrasolar planets. At the
time these were state-of-the-art echelle spectrometers. Figure \ref{fig:PSF}
illustrates the progress in echelle spectrometer design over the past
generation. UVES \citep{Dekker_2000} is arguably the first modern precision 
RV spectrometer capable of producing 1 \ms  uncertainty \citep{Butler_2019}.
Subsequent echelle spectrometers built specifically for precision RV
measurements have R~$>$120K (HARPS, PFS, ESPRESSO).

\begin{figure}
\centering
\includegraphics[scale=.5,angle=180]{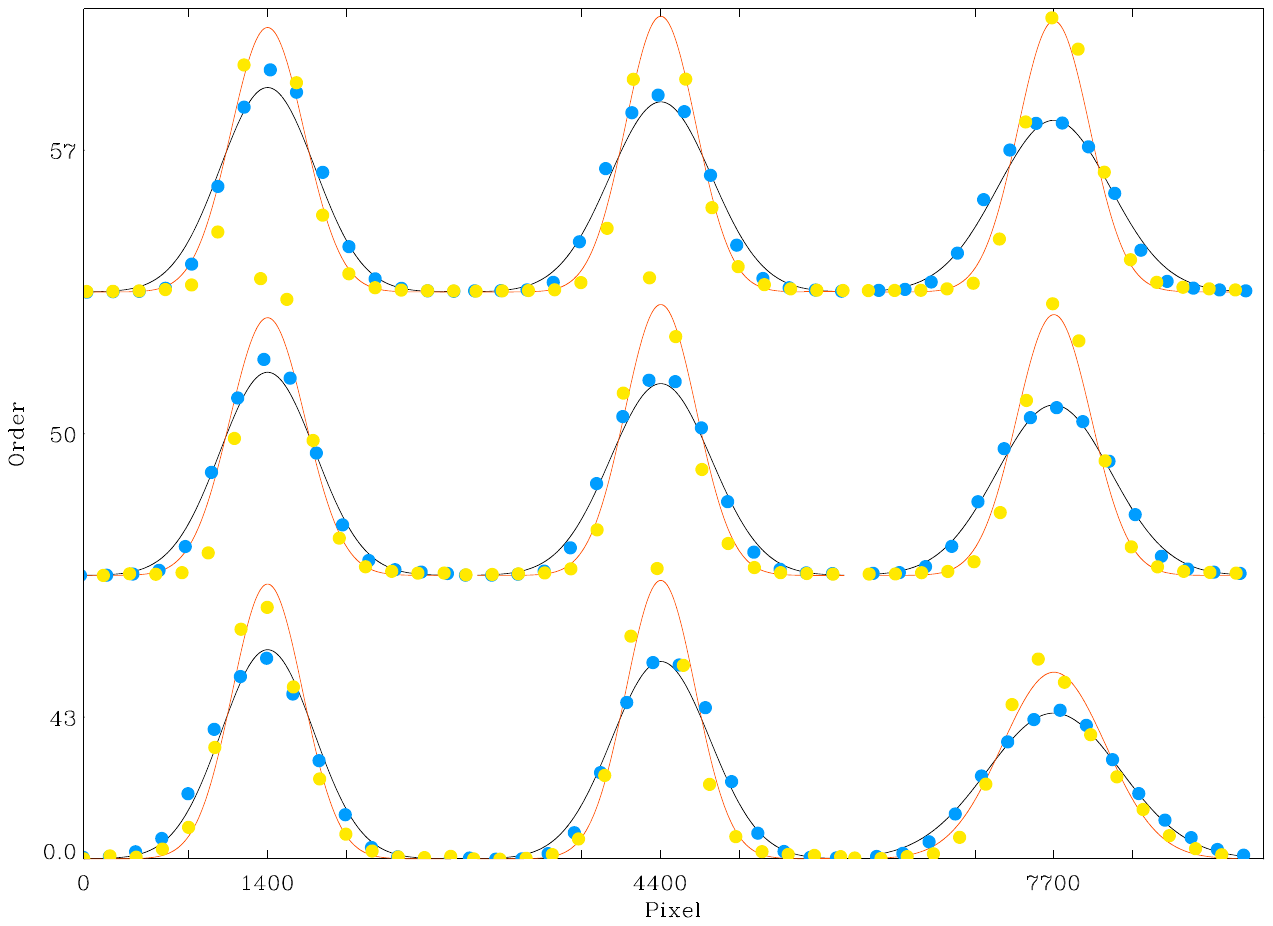}
\caption{The PSF of ESPRESSO over the iodine format. Order 43 corresponds to
(vacuum) wavelengths in 5160~\AA\ to 5217~\AA\ region. Order 50 covers the 5432
to 5505~\AA\ region. Order 57 covers 5794~\AA\ to 5872~\AA. The central pixels (1400,
4400, 7700) cover the left, center, and right of the CCD format. The ticks on
either side of the pixel indicate +/-3 pixels. Black is the PSF derived from
iodine in the HR21 mode (R~136K). Red is the PSF in the UHR mode (R~194K). The
blue and orange dots are the LFC for the HR21 and UHR modes
respectively.\label{fig:PSF_ESPRESSO}}
\end{figure}

As with all echelle spectrometers, the ESPRESSO PSF varies over the echelle
format. Figure \ref{fig:PSF_ESPRESSO} shows the ESPRESSO PSF derived from the
iodine for the UHR mode (red) and HR21 mode (black) modes. The width of the PSF
is largest at the far right of the format at the lowest orders. The blue and
orange dots are the LFC for HR21 and UHR respectively. The PSFs derived from the
LFC lines and the iodine lines show the same trends, especially the increase in
width at the bottom right of the echelle format.

\subsection{ESPRESSO Doppler velocities} \label{sec:velocities}

Commissioning runs are always stress filled. Equipment built and tested in the
lab typically does not initially work in the harsher mountain environment. The
new equipment must then be accurately mounted and aligned with the existing
system. Along with delivering a working instrument, the primary goal of a
commissioning run is to demonstrate that the new instrument is promising,
capable of providing data at a world class level, and thus justifying further
time and resources into the development of the instrument and data reduction.
For a precision RV spectrometer, this requires observing standard stable
stars.

The ESPRESSO iodine commissioning run was especially difficult as the program
was only allocated 40 minutes of evening and morning twilight to carry out the
observations. As every experienced observer is acutely aware, the first hour of
any night is often the most challenging. In addition to the the stable test
stars observed through the iodine cell, we also had to fit in ``template
spectra'' of the stars, taken without iodine, and calibration spectra of
essentially featureless rapidly rotating B-stars during the brief 40 minute
observing intervals. As this was a commissioning run, the iodine cell was not
automated to move in and out of the beam. This necessitated running several
hundred meters across the mountain from the observing room to the spectrometer
room whenever the iodine cell needed to be moved in or out of the beam. Under
these constraints, we were only able to observe five stars, listed in Table
\ref{tab:stars}.

\begin{table}
%\centering

\caption{Stable stars observed with the ESPRESSO iodine cell} 

\label{tab:stars}
\begin{tabular}{lllllllllll}
\tableline 
Star      & V    & B-V  & Total & Number of & Obs. per & Time per  & Full time    & RV & Internal    \\
          & Mag  &      & Obs.  & Nights    & night    & obs.      & span/night   & RMS      & Uncertainty \\
          &      &      &       &           &          & (s)       & (min)        & (\ms)    & (\ms)       \\
\tableline                                                                                 
HD53706   & 6.85 & 0.78 & 12    & 4         &    3     & 150 - 200 &  8.8 - 11.9  & 0.04     & 0.21        \\
HD59468   & 6.71 & 0.71 & 26    & 6         &   3-5    &  90 - 200 &  6.8 - 14.4  & 0.79     & 0.21        \\
HD102365  & 4.88 & 0.67 & 23    & 6         &   3-8    &  70 - 100 &  5.2 - 17.1  & 1.94     & 0.39        \\
HD161612  & 7.20 & 0.71 & 12    & 4         &   3      &  200      & 11.6 - 11.7  & 0.96     & 0.31        \\
HD199190  & 6.87 & 0.62 & 12    & 4         &   3      &  150      &  8.3 -  9.3  & 1.18     & 0.40        \\
\tableline                                 
\end{tabular}
\end{table}

These stars were selected from a list of the brightest stable stars in the
Magellan PFS survey in the limited right ascension band available during evening
and morning twilight. These stars have been observed with Magellan PFS for 12 to
15 years. With the exception of HD102365, the PFS RV RMS of these stars
are all $<$ 2 \ms. They span the range from early G to early K dwarfs.

We typically took three or four consecutive observations on each star, with a
S/N of $\approx$100 for the individual observations. Taking multiple observations
spanning 10-to-15 minutes allows averaging over the p-mode oscillations. 
The gold standard for minimizing jitter due to p-modes and granulation is 
observing a star three times a night separated by an hour or two, with at least 
10 minutes of dwell time for each set \cite{Pepe_2011}.  Due to 
the tight constraint of 40 minutes for each observing session, this was not 
possible on this commissioning run.  We were only able to take one set of 
observations for each star with a dwell time of 5 to 17 minutes, so our results 
do not average over the granulation, and do not fully average over the p-modes.

The nightly binned RV RMS ranges from 0.04 to 1.94 \ms. Observations taken
over a short time span typically show a smaller RV RMS that a decade long
string. On short timescales instrumental and stellar jitter variations are
minimized.

\begin{figure}
\centering
\includegraphics[scale=.5]{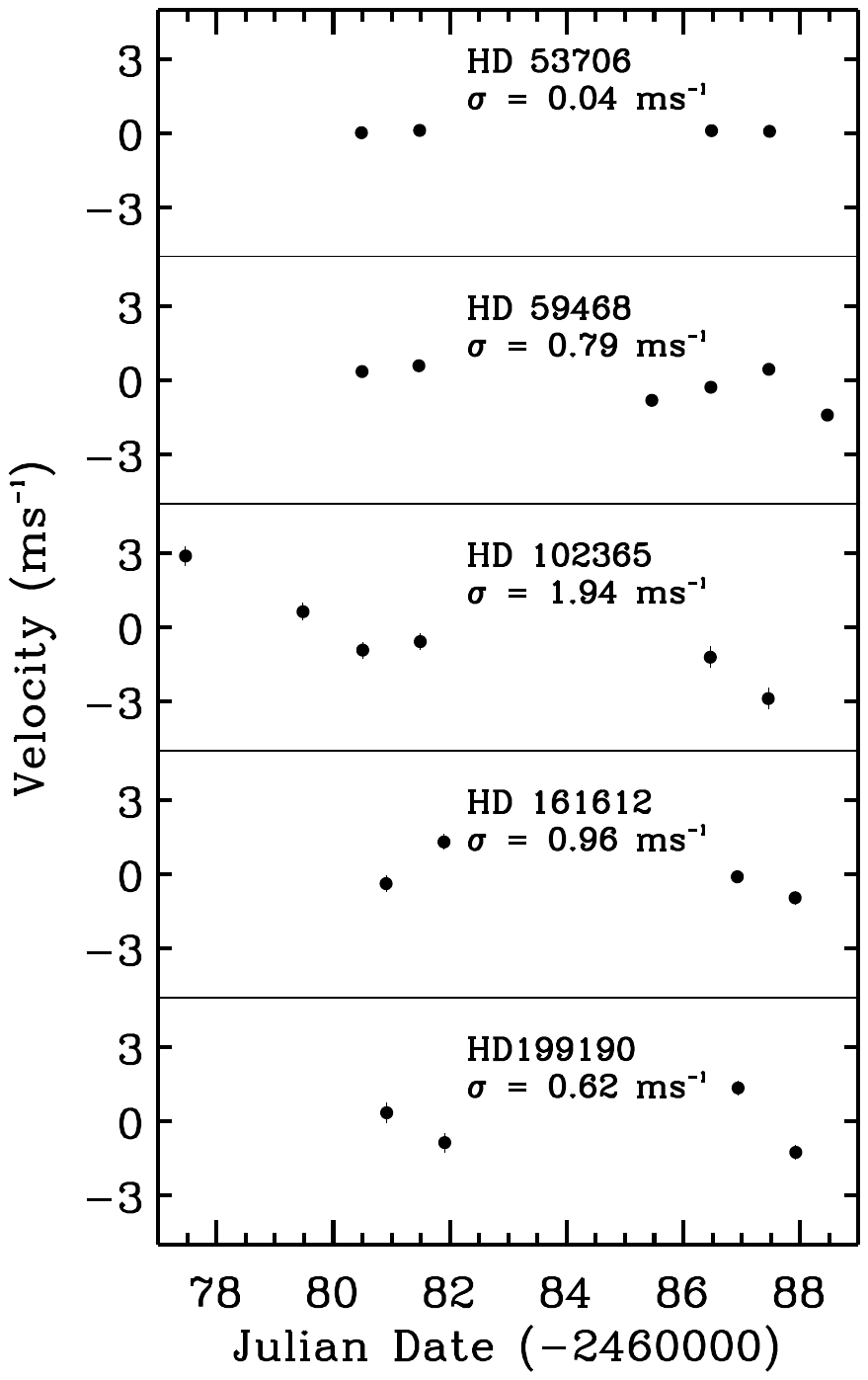}
\caption{Nightly binned RVs of the five ESPRESSO test stars.  
The nightly binned RV RMS is also shown.  The absurdly good RMS for
HD53706 is due to small number statistics.
\label{fig:radial_vel}}
\end{figure}

The resulting nightly binned RVs are shown in Figure
\ref{fig:radial_vel}, with the RV RMS and internal uncertainties given in
Table \ref{tab:stars}. The resulting nightly binned internal uncertainty of the
stars range from 0.21 to 0.40 \ms. As expected, the late G and early K stars
have a smaller internal uncertainty and RV RMS relative to the earlier G
stars. The later stars have more and deeper lines, and hence more RV
information.

\section{Reference Spectra} \label{sec:refspectra}

A 1 \ms Doppler shift on current state-of-the-art spectrometers amounts to
roughly 40 silicon atoms on the CCD substrate. At 2 \cms\ the error budget drops
to 1 silicon atom. At this level any microscopic change to the instrument must
be accounted for. Precision RV spectrometers thus require a reference
spectrum cable of tracking changes on atomic length scales.

At a minimum the reference spectrum sets the wavelength scale. Velocity
precision is directly tied to the uncertainty of the wavelength scale. An ideal
reference spectrum has a high density of extremely stable and narrow (delta
function) lines. The information density of the reference must be significantly
higher than the Doppler information in the target.

\begin{figure}
\centering
\includegraphics[scale=.5]{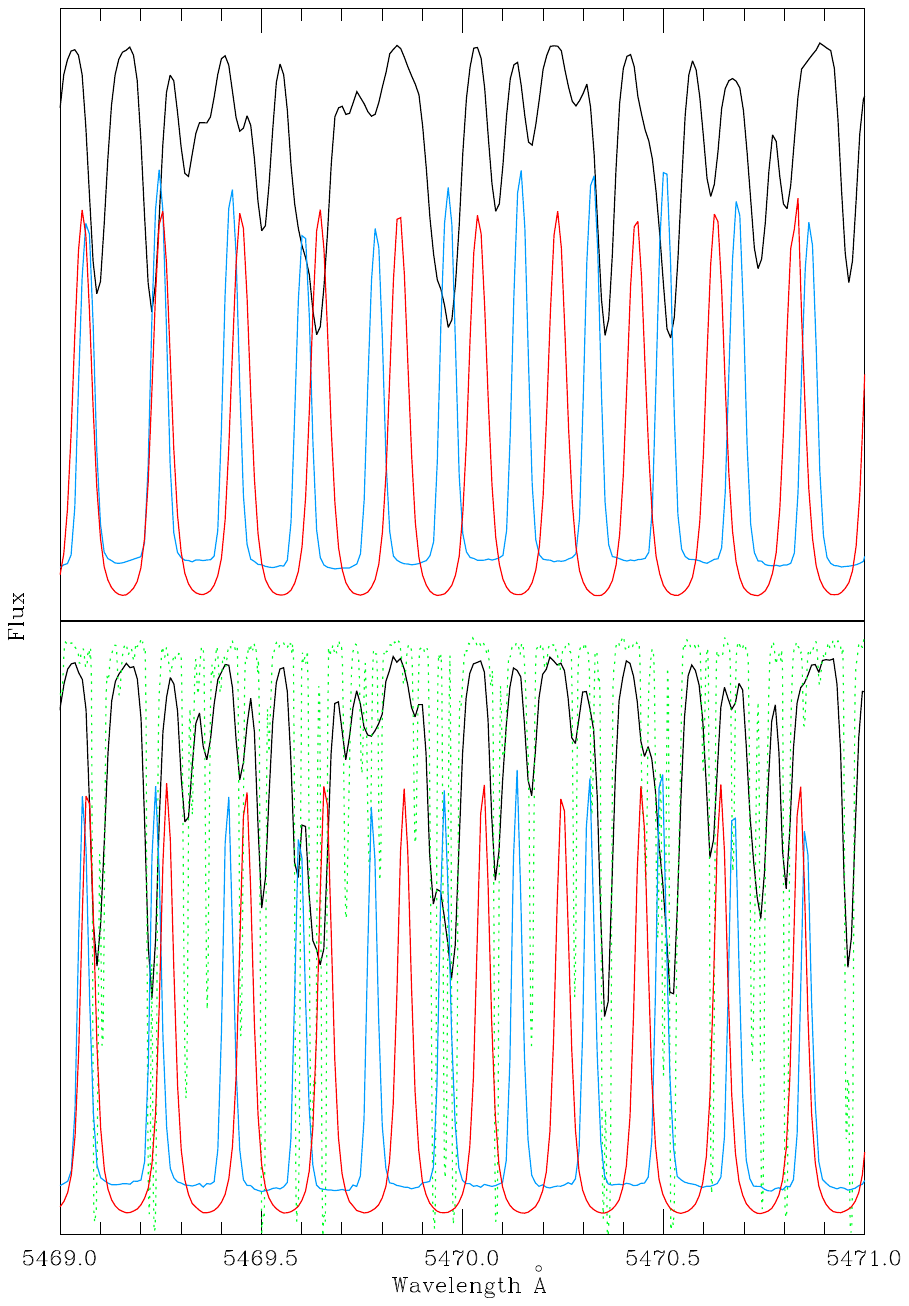}
\caption{ESPRESSO reference spectra in the HR21 observing mode (top) and in the 
UHR mode (bottom) over 2~\AA\ in the middle of the iodine region.  Red is the 
Fabry-P\'{e}rot.  Blue is LFC.  Black is the iodine. The green dots are the 
underlying iodine spectrum at a resolution of 2 million from the NIST FTS. 
\label{fig:ref_espr}}
\end{figure}

Figure \ref{fig:ref_espr} shows three of the reference sources for ESPRESSO over
a 2~\AA\ region in the middle of the iodine region. The Th/Ar HCL references is
not included in these plots because over a typical 2~\AA\ region there are only
zero, one or two lines from this lamp. Over this 2~\AA\ region, there are 10
Fabry-P\'{e}rot lines and 11 LFC lines. Due to blending, the number of iodine
lines is a function of the resolving power. In the HR21 observing mode there are
about 17 iodine lines. In the UHR mode there are roughly 25 iodine lines. At the
resolving power of the NIST FTS ($\sim$2 million), there are roughly 30 lines.

\begin{figure}
\centering
\includegraphics[scale=.5,angle=180]{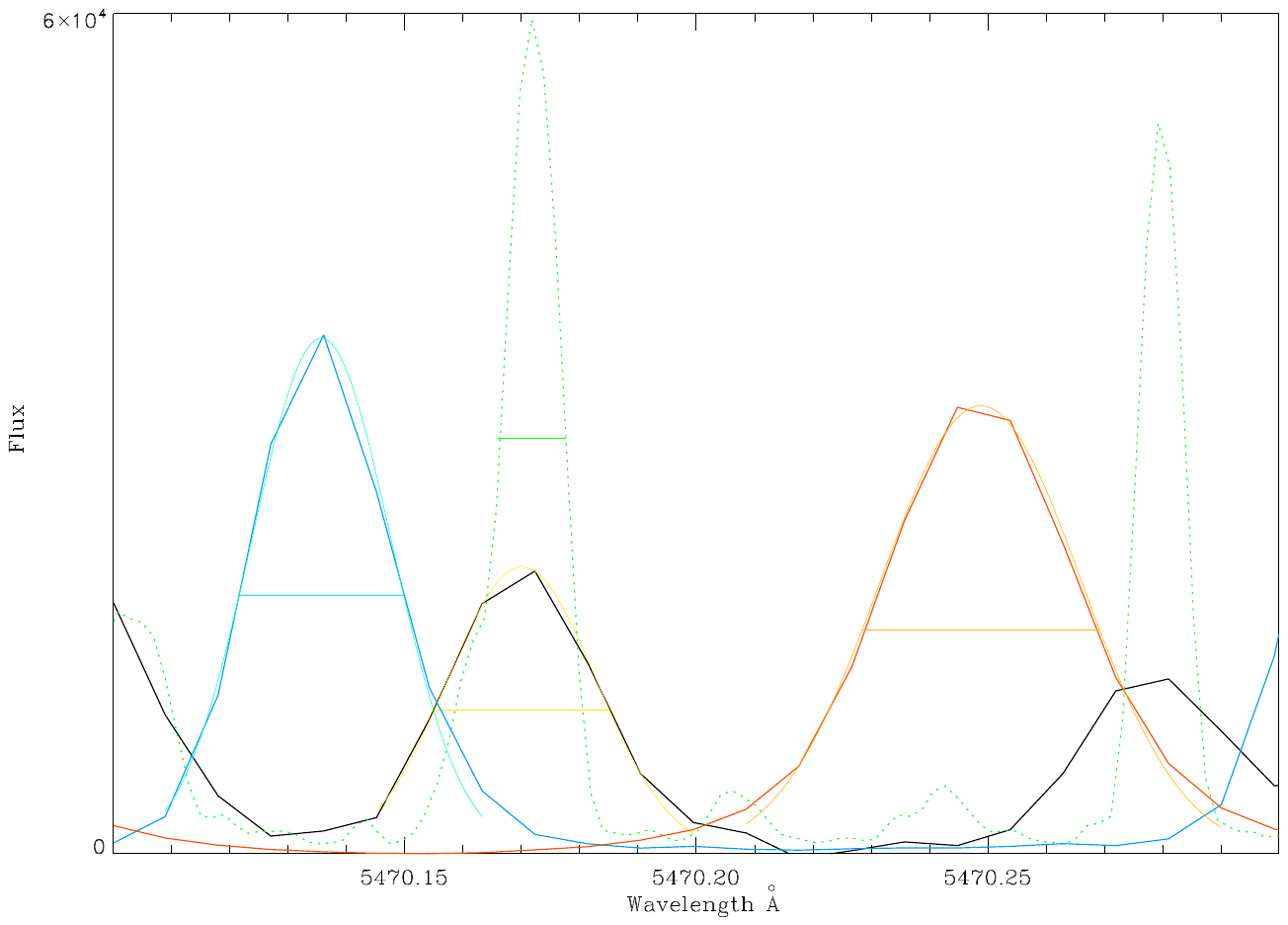}
\caption{ESPRESSO reference spectra in the UHR mode over
0.3~\AA .  Red is the Fabry-P\'{e}rot.  Blue is the LFC.  Black is the iodine.
The green dots are the underlying iodine spectrum at a resolving power of 2 
million from the
NIST FTS. Highly over-sampled Gaussians are fit to each of the three reference 
lines. The FWHM of each reference line is shown. \label{fig:UHR_comparison} }
\end{figure}

The Fabry-P\'{e}rot and LFC are emission spectra. Iodine and stellar spectra are
absorption spectra. It is instructive to treat the all the reference spectra as
if they were emission spectra. Figure \ref{fig:UHR_comparison} shows the three
reference spectra in the UHR mode with the continuum subtracted off. The iodine
spectrum is inverted (multiplied by -1) after the continuum subtraction to
generate a pseudo emission spectrum. At the resolution and sampling of ESPRESSO,
most of the underlying iodine lines (green dots from the NIST FTS atlas) are
blended. The iodine line at 5470.17~\AA\ is relatively unblended. Highly
over-sampled Gaussian fits for the iodine line (5470.17~\AA ), the laser line
(5470.13~\AA ) and the Fabry-P\'{e}rot line (5470.25~\AA ) are shown. The full
width at half maximum (FWHM) are listed in Table \ref{Tab:widths}.

\begin{table}
%\centering
\caption{Width of Reference Lines near 5470.17~\AA\ for ESPRESSO UHR}

\label{Tab:widths}
\begin{tabular}{llllllll}
\tableline
Source & FWHM (\AA) & Wavelength/FWHM \\
   \tableline                                                
NIST FTS I2  & 0.0115 & 417289 \\
ESPRESSO I2  & 0.0301 & 181974 \\
LFC          & 0.0283 & 193352 \\
Fabry-P\'{e}rot  & 0.0407 & 134387 \\
\tableline    
\end{tabular}
\end{table}

The ``Wavelength/FWHM'' of the LFC line is consistent with the resolution
derived from iodine for the ESPRESSO UHR mode. This suggests that the laser
lines are approaching  effectively delta functions at the resolution of the
UHR mode. The intrinsic width of the underlying iodine line from the NIST FTS
(green dots) is roughly two times narrower than the resolution of ESPRESSO in
UHR mode. When convolved with the ESPRESSO PSF in UHR mode, the resulting iodine
line is 6\% wider than the laser. The FWHM of Fabry-P\'{e}rot line is 44\% wider
than the laser.

\section{Summary/Conclusion} \label{sec:summary}

Achieving an uncertainty in the RV below 1~\ms in the long term is
extremely difficult. If a long term uncertainty of a few \cms were
achievable, it would be possible to detect earth-analogs orbiting nearby stars,
and directly measure the expansion of the Universe with the Sandage test if
maintained over a period of decades. Finding earth-analogs is especially
fraught. The Earth induces a RV semi-amplitude of 8 \cms on the Sun. Even
slowly rotating stars jitter at the 0.5 to 1 \ms level. Finding earth-analogs
with precision RVs requires improving the intrinsic  uncertainty of
Doppler RV spectrometers to the 2 \cms level, and accurately accounting
for and modeling the stochastic stellar jitter which is operating on all time
scales. There are four main sources for stellar jitter. For sun-like stars,
pressure mode (p-mode) oscillations have a timescale of minutes. Photospheric
granulation has a timescale of minutes to a few days. Rotationally modulated
spots and faculae have a timescale of weeks to months. Stellar magnetic cycles
have a period of years to decades \citep{McWilliam_2026}. All of these can
produce pseudo Doppler shifts of order 1 \ms.

One of the goals of the ANDES spectrograph planned for the ESO
ELT is to directly measure the expansion of the Universe using the Sandage test
\citep{Sandage_1962}, and requires maintaining this 2~\cms level over a period
of one to two decades. This places even more stringent demands on the wavelength
calibration.

Of these two problems -- intrinsic instrumental  uncertainty and stellar
jitter -- the first is perhaps more tractable, though still extremely difficult.
The critical issues in improving the uncertainty of a Doppler spectrometer
are resolution, sampling, and the reference spectrum.

The three reference spectra available on ESPRESSO each have advantages and
disadvantages. The LFC produces a high density of narrow lines over the full
wavelength range of the spectrograph. But the reference spectra are not taken
simultaneously with the target observation, so they can not track any changes in
the spectrometer between the time of the target and reference observations.
After several years of development, the LFC is still not used as the primary
reference for ESPRESSO precision RV measurements. Given the intrinsic
narrowness of the laser lines, they could possibly be used to determine and
track changes in the spectrometer PSF. It would be helpful to scan the ESPRESSO
LFC with an FTS at much better resolution in order to model the ESPRESSO PSF on
a sub-pixel scale and account for any very small intrinsic width of the laser
lines, in a similar way to studies of two infrared LFCs by
\cite{Ycas_2010}, Fig. 7 and \cite{Reiners_2024}. In both of these studies, the
lineshapes were similar to the instrumental lineshape of the FTS, suggesting
that higher resolution would be of benefit. \cite{Ycas_2010} gave a peak S/N of
4000 at a resolving power of around 2.5$\times$10$^6$ in the wavelength range
1.3 - 1.9 $\mu$m. \cite{Reiners_2024} gave a peak S/N of about 400, with a
resolving power of 2.3$\times$10$^6$ in the wavelength region 7000 - 9000~\AA. An
improvement of at least an order of magnitude in the S/N would be needed to
observe any possible side-modes of the LFC comb lines.

The Fabry-P\'{e}rot is currently the default reference for ESPRESSO precision
RV observations. Relative to the LFC, the Fabry-P\'{e}rot reference has
the advantage of being taken simultaneously with the target observation. At the
resolution of ESPRESSO, the Fabry-P\'{e}rot lines have significant width.
Between the width and density of the Fabry-P\'{e}rot lines, they carry somewhat
less information than the laser or iodine lines.

The LFC and the Fabry-P\'{e}rot produce emission lines, and they are carried on
a separate fiber from the target source. Therefore they sample the spectrometer
optics slightly differently relative to the target. They are both subject to
systematic errors due to motion on atomic scales between the target and
reference fibers. Maintaining the spacing between the fibers at the level of a
single atom in the face of earthquakes and other mechanical vibrations is
difficult and expensive. The intrinsic widths of the iodine lines are much
smaller than the ESPRESSO UHR mode, and their observed widths in this mode are
nearly as narrow as those from the LFC. The density of the iodine lines is
higher than the LFC. The iodine lines are in absorption, as are precision
RV stellar targets, and lines from quasar absorption spectra used to place
constraints on a possible variation in the fine-structure constant in the early
Universe. The iodine reference is carried on the beam of starlight, so the
reference is both both simultaneous and spatially invariant relative to the
target. Even during an earthquake, iodine provides an invariant reference.

Achieving  a low short-term uncertainty is much easier than long term. This
could be called the Paul Simon law: Everything put together falls apart. Dating
back to HARPS (\citep{Rupprect_2004}), super-stabilized spectrometers have an
impressive record of producing  uncertainties at the 1 \ms or better level.
The problem is that super-stabilized spectrometers cannot  easily recover
from changes to the system. Any successful precision RV spectrometer that
survives long enough will have optical and detector upgrades. These will change
the spectrometer PSF. The 2015 upgrade to HARPS led to a $\approx$ 8~\ms 
RV offset (\citep{Trifonov_2020}, Figure 9). Reference sources also age over
decades, but while the LFC, Th/Ar HCLs, and Fabry-P\'{e}rot sources will require
replacement of key parts, a well-built and maintained iodine cell should last
through many decades, provided it is not broken or heated above 100~\degC\ for
many hours. This  has been demonstrated with the Keck/HIRES iodine cell,
currently 30+ years old, and the VLT2/UVES cell, which is 25 years old.

Achieving {a long-term uncertainty of a few \cms}, if it is possible, is going
to be very difficult. Solving this problem is going to require everything we can
throw at it. The obvious improvements are observing at a higher resolving power
with better sampling. Current state-of-the-art spectrometers have 9-to-10~$\mu$m
pixels. Next generation CMOS detectors with 3~$\mu$m pixels should be available
in the next few years. The development of reference spectra must also continue.
Reference spectra, including Fabry-P\'{e}rot, LFCs, and iodine cells, all
benefit from being scanned by FTS at a resolving power of several million. 
We note that two FTS instruments have historically been used to measure
calibration spectra of iodine cells and Th/Ar HCLs for ground-based astronomical
spectrographs - the 1-m FTS at the McMath Solar Observatory and the FTS at NIST
used in the current work. The first was decomissoned in 2012, and access to the
FTS at NIST is no longer guaranteed, following the move of the NIST Atomic
Spectroscopy Group to NASA Goddard Space Flight Center. Maintaining the ability
to scan calibration sources at high resolution will be essential for future
astronomical spectrographs.

Along with Fabry-P\'{e}rot and LFCs, iodine cells should be part of the mix.
Iodine has two significant advantages. It is an absorption spectrum carried
directly on the beam of starlight. It is thus identically sampled both
temporally and spatially with the target. Iodine is the only current technique
that can both provide a wavelength scale and a direct measurement of the
instrumental PSF through the same light path as the astrophysical spectrum. In
Table \ref{tab:stars}, we show that the iodine cell technique is capable of
measuring RVs with an internal uncertainty as low as 0.21~\ms, similar to the
value of 0.25~\ms demonstrated in the commissioning of ESPRESSO
\citep{Pepe_2021}. The principal contributors to this uncertainty are the
stellar jitter and the photon noise from the short length of time that was
available for the observations. The principal disadvantages of the iodine cell
technique are that the iodine lines are limited to the region 5000~\AA\ to
6200~\AA, and that by superimposing the iodine lines on the stellar spectrum,
features in the spectrum may be obscured and the resulting data analysis is more
complex. By combining all three reference spectra, the iodine cell can track the
overall changes in the instrumental PSF, while the LFC and the Fabry-P\'{e}rot
can provide the wavelength scale over a larger wavelength region.

For the ESPRESSO iodine cell to come to full fruition, a computer controlled
mechanism to move the iodine cell in and out of the beam must be constructed.
Time needs to be allocated to observing a handful of stable stars with the
iodine cell over an extended period to allow for the development of an improved
ESPRESSO specific iodine RV reduction package. We are especially intrigued
by the possibility of observing known stable stars with ESPRESSO in UHR mode.
Precision RV measurements have not yet been taken in this resolution
regime.

\begin{acknowledgements}
We thank Darren Dougan of the Big Questions Institute, Sydney, for funding the 
construction of the iodine cell and for funding Paul Butler's travel 
to Paranal. We thank Gaspare LoCurto for arranging and assisting with the 
installation of the iodine cells, and for writing custom observation scripts for 
the observations. We thank the ESPRESSO instrument scientists and telescope 
operators for their support and assistance during the observations. All data 
analyzed in this work are available from the ESO archives under the Program ID 
60.A-9680(A). 
\end{acknowledgements}
\vspace{5mm}
\facility{VLT(ESPRESSO)}

\bibliography{I2_ESPRESSO_paper.bib}{}

@ARTICLE{Griffin_1987,
       author = {{Griffin}, R.F. and {Griffin}, R.E.},
        title = "{On the possibility of determining stellar radial velocities to 0.01 km s$^{-1}$.}",
      journal = {\mnras},
         year = 1973,
        month = jan,
       volume = {162},
        pages = {243-253},
          doi = {10.1093/mnras/162.3.243},
       adsurl = {https://ui.adsabs.harvard.edu/abs/1973MNRAS.162..243G}
}

@ARTICLE{Campbell_1979,
       author = {{Campbell}, B. and {Walker}, G.~A.~H.},
        title = "{Precision radial velocities with an absorption cell.}",
      journal = {\pasp},
         year = 1979,
        month = aug,
       volume = {91},
        pages = {540-545},
          doi = {10.1086/130535},
       adsurl = {https://ui.adsabs.harvard.edu/abs/1979PASP...91..540C}
}

@article{Butler_2019,
doi = {10.3847/1538-3881/ab4905},
url = {https://doi.org/10.3847/1538-3881/ab4905},
year = {2019},
month = {dec},
publisher = {The American Astronomical Society},
volume = {158},
number = {6},
pages = {251},
author = {Butler, R. P. and Jones, H. R. A. and Feng, F. and Tuomi, M. and Anglada-Escudé, G. and Keiser, Sandy},
title = {A Reanalysis of the UVES M Dwarf Planet Search Program*},
journal = {The Astronomical Journal},
}

@ARTICLE{MarcyButler_1992,
       author = {{Marcy}, Geoffrey W. and {Butler}, R.~P.},
        title = "{Precision Radial Velocities with an Iodine Absorption cell}",
      journal = {\pasp},
         year = 1992,
        month = apr,
       volume = {104},
        pages = {270},
          doi = {10.1086/132989},
       adsurl = {https://ui.adsabs.harvard.edu/abs/1992PASP..104..270M}
}

@ARTICLE{MarcyButler_1996,
       author = {{Marcy}, Geoffrey W. and {Butler}, R. Paul},
        title = "{A Planetary Companion to 70 Virginis}",
      journal = {\apjl},
         year = 1996,
        month = jun,
       volume = {464},
        pages = {L147},
          doi = {10.1086/310096},
       adsurl = {https://ui.adsabs.harvard.edu/abs/1996ApJ...464L.147M}
}

@ARTICLE{Butler_1997,
       author = {{Butler}, R. Paul and {Marcy}, Geoffrey W. and {Williams}, Eric and {Hauser}, Heather and {Shirts}, Phil},
        title = "{Three New ``51 Pegasi-Type'' Planets}",
      journal = {\apjl},
         year = 1997,
        month = jan,
       volume = {474},
       number = {2},
        pages = {L115-L118},
          doi = {10.1086/310444},
       adsurl = {https://ui.adsabs.harvard.edu/abs/1997ApJ...474L.115B}
}

@ARTICLE{Butler_1996b,
       author = {{Butler}, R. Paul and {Marcy}, Geoffrey W.},
        title = "{A Planet Orbiting 47 Ursae Majoris}",
      journal = {\apjl},
         year = 1996,
        month = jun,
       volume = {464},
        pages = {L153},
          doi = {10.1086/310102},
       adsurl = {https://ui.adsabs.harvard.edu/abs/1996ApJ...464L.153B}
}

@ARTICLE{Butler_1996,
       author = {{Butler}, R.~P. and {Marcy}, G.~W. and {Williams}, E. and {McCarthy}, C. and {Dosanjh}, P. and {Vogt}, S.~S.},
        title = "{Attaining Doppler Precision of 3 M s-1}",
      journal = {\pasp},
         year = 1996,
        month = jun,
       volume = {108},
        pages = {500},
          doi = {10.1086/133755},
       adsurl = {https://ui.adsabs.harvard.edu/abs/1996PASP..108..500B}
}

@inproceedings{Crause_18,
author = {Lisa A. Crause and R. Paul Butler and Gillian Nave and Rudi Kuhn and Blaine Lomberg and Alexei Kniazev and Steven M. Crawford and {\'E}ric Depagne},
title = {{Commissioning the SALT High Resolution Spectrograph’s iodine cell}},
volume = {10702},
booktitle = {Ground-based and Airborne Instrumentation for Astronomy VII},
editor = {Christopher J. Evans and Luc Simard and Hideki Takami},
organization = {International Society for Optics and Photonics},
publisher = {SPIE},
pages = {107025S},
year = {2018},
doi = {10.1117/12.2307195},
URL = {https://doi.org/10.1117/12.2307195}
}

@mastersthesis{Butler_1987,
author = {R. Paul Butler},
school = {San Francisco State University},
year = {1987}
}

@ARTICLE{Mayor_1995,
       author = {{Mayor}, Michel and {Queloz}, Didier},
        title = "{A Jupiter-mass companion to a solar-type star}",
      journal = {\nat},
         year = 1995,
        month = nov,
       volume = {378},
       number = {6555},
        pages = {355-359},
          doi = {10.1038/378355a0},
       adsurl = {https://ui.adsabs.harvard.edu/abs/1995Natur.378..355M}
}

@ARTICLE{Cochran_1988,
       author = {{Cochran}, William D.},
        title = "{Confirmation of Radial Velocity Variability in Arcturus}",
      journal = {\apj},
         year = 1988,
        month = nov,
       volume = {334},
        pages = {349},
          doi = {10.1086/166841},
       adsurl = {https://ui.adsabs.harvard.edu/abs/1988ApJ...334..349C}
}

@INPROCEEDINGS{Cochran_1985,
       author = {{Cochran}, William D. and {Young}, Brenda W.},
        title = "{The Mcdonald Observatory High Precision Radial Velocity Spectrometer}",
    booktitle = {Stellar Radial Velocities},
         year = 1985,
       editor = {{Philip}, A.~G.~D. and {Latham}, David W.},
        month = jan,
        pages = {109-120},
       adsurl = {https://ui.adsabs.harvard.edu/abs/1985srv..conf..109C}
}

@ARTICLE{Smith_1982,
       author = {{Smith}, M.~A.},
        title = "{Precise radial velocities. I. A preliminary search for oscillations in Arcturus.}",
      journal = {\apj},
         year = 1982,
        month = feb,
       volume = {253},
        pages = {727-734},
          doi = {10.1086/159673},
       adsurl = {https://ui.adsabs.harvard.edu/abs/1982ApJ...253..727S}
}

@INPROCEEDINGS{Merline_1985,
       author = {{Merline}, W.~J.},
        title = "{Radial Velocity Information in Solar-Type Spectra}",
    booktitle = {Stellar Radial Velocities},
         year = 1985,
       editor = {{Philip}, A.~G.~D. and {Latham}, David W.},
        month = jan,
        pages = {87},
       adsurl = {https://ui.adsabs.harvard.edu/abs/1985srv..conf...87M}
}

@ARTICLE{Smith_1987,
       author = {{Smith}, P.~H. and {McMillan}, R.~S. and {Merline}, W.~J.},
        title = "{Evidence for Periodic Radial Velocity Variations in Arcturus}",
      journal = {\apjl},
         year = 1987,
        month = jun,
       volume = {317},
        pages = {L79},
          doi = {10.1086/184916},
       adsurl = {https://ui.adsabs.harvard.edu/abs/1987ApJ...317L..79S}
}

@INPROCEEDINGS{McMillan_1986,
       author = {{McMillan}, R.~S. and {Smith}, P.~H. and {Frecker}, J.~E. and {Merline}, W.~J. and {Perry}, M.~L.},
        title = "{A Fabry-Perot interferometer for accurate measurement of temporal changes in stellar Doppler shift.}",
    booktitle = {Instrumentation in astronomy VI},
         year = 1986,
       editor = {{Crawford}, David L.},
       series = {Society of Photo-Optical Instrumentation Engineers (SPIE) Conference Series},
       volume = {627},
        month = jan,
        pages = {2-19},
          doi = {10.1117/12.968068},
       adsurl = {https://ui.adsabs.harvard.edu/abs/1986SPIE..627....2M}
}

@ARTICLE{Butler_2006,
       author = {{Butler}, R.~P. and {Wright}, J.~T. and {Marcy}, G.~W. and {Fischer}, D.~A. and {Vogt}, S.~S. and {Tinney}, C.~G. and {Jones}, H.~R.~A. and {Carter}, B.~D. and {Johnson}, J.~A. and {McCarthy}, C. and {Penny}, A.~J.},
        title = "{Catalog of Nearby Exoplanets}",
      journal = {\apj},
         year = 2006,
        month = jul,
       volume = {646},
       number = {1},
        pages = {505-522},
          doi = {10.1086/504701},
archivePrefix = {arXiv},
       eprint = {astro-ph/0607493},
 primaryClass = {astro-ph},
       adsurl = {https://ui.adsabs.harvard.edu/abs/2006ApJ...646..505B}
}

@ARTICLE{Trifonov_2020,
       author = {{Trifonov}, Trifon and {Tal-Or}, Lev and {Zechmeister}, Mathias and {Kaminski}, Adrian and {Zucker}, Shay and {Mazeh}, Tsevi},
        title = "{Public HARPS radial velocity database corrected for systematic errors}",
      journal = {\aap},
         year = 2020,
        month = apr,
       volume = {636},
          eid = {A74},
        pages = {A74},
          doi = {10.1051/0004-6361/201936686},
archivePrefix = {arXiv},
       eprint = {2001.05942},
 primaryClass = {astro-ph.EP},
       adsurl = {https://ui.adsabs.harvard.edu/abs/2020A&A...636A..74T}
}

@inproceedings{Rupprect_2004,
author = {Gero Rupprecht and Francesco Pepe and Michel Mayor and Didier Queloz and Francois Bouchy and Gerardo Avila and Willy Benz and Jean-Loup Bertaux and X. Bonfils and Th. Dall and Bernard Delabre and Hans Dekker and Wolfgang Eckert and Michel Fleury and Alain Gilliotte and Domingo Gojak and Juan Carlos Guzman and Dominique Kohler and Jean-Louis Lizon and G. Lo Curto and Antonio Longinotti and Christophe Lovis and Denis Megevand and Luca Pasquini and Javier Reyes and Jean-Pierre Sivan and Danuta Sosnowska and R. Soto and Stephane Udry and Arno Van Kesteren and Luc Weber and Ueli Weilenmann},
title = {{The exoplanet hunter HARPS: performance and first results}},
volume = {5492},
booktitle = {Ground-based Instrumentation for Astronomy},
editor = {Alan F. M. Moorwood and Masanori Iye},
organization = {International Society for Optics and Photonics},
publisher = {SPIE},
pages = {148 -- 159},
year = {2004},
doi = {10.1117/12.551267},
URL = {https://doi.org/10.1117/12.551267}
}

@ARTICLE{Sandage_1962,
       author = {{Sandage}, Allan},
        title = "{The Change of Redshift and Apparent Luminosity of Galaxies due to the Deceleration of Selected Expanding Universes.}",
      journal = {\apj},
         year = 1962,
        month = sep,
       volume = {136},
        pages = {319},
          doi = {10.1086/147385},
       adsurl = {https://ui.adsabs.harvard.edu/abs/1962ApJ...136..319S}
}

@INPROCEEDINGS{ANDES,
       author = {{Marconi}, A. and {Abreu}, M. and {Adibekyan}, V. and {Alberti}, V. and {Albrecht}, S. and {Alcaniz}, J. and {Aliverti}, M. and {Allende Prieto}, C. and {Alvarado-Gomez}, J.~D. and {Alves}, C.~S. and {Amado}, P.~J. and {Amate}, M. and {Andersen}, M.~I. and {Antoniucci}, S. and {Artigau}, E. and {Bailet}, C. and {Baker}, C. and {Baldini}, V. and {Balestra}, A. and {Barnes}, S.~A. and {Baron}, F. and {Barros}, S.~C.~C. and {Bauer}, S.~M. and {Beaulieu}, M. and {Bellido-Tirado}, O. and {Benneke}, B. and {Bensby}, T. and {Bergin}, E.~A. and {Berio}, P. and {Biazzo}, K. and {Bigot}, L. and {Bik}, A. and {Birkby}, J.~L. and {Blind}, N. and {Boebion}, O. and {Boisse}, I. and {Bolmont}, E. and {Bolton}, J.~S. and {Bonaglia}, M. and {Bonfils}, X. and {Bonhomme}, L. and {Borsa}, F. and {Bouret}, J. -C. and {Brandeker}, A. and {Brandner}, W. and {Broeg}, C.~H. and {Brogi}, M. and {Brousseau}, D. and {Brucalassi}, A. and {Brynnel}, J. and {Buchhave}, L.~A. and {Buscher}, D.~F. and {Cabona}, L. and {Cabral}, A. and {Calderone}, G. and {Calvo-Ortega}, R. and {Cantalloube}, F. and {Canto Martins}, B.~L. and {Carbonaro}, L. and {Caujolle}, Y. and {Chauvin}, G. and {Chazelas}, B. and {Cheffot}, A. -L. and {Cheng}, Y.~S. and {Chiavassa}, A. and {Christensen}, L. and {Cirami}, R. and {Cirasuolo}, M. and {Cook}, N.~J. and {Cooke}, R.~J. and {Coretti}, I. and {Covino}, S. and {Cowan}, N. and {Cresci}, G. and {Cristiani}, S. and {Cunha Parro}, V. and {Cupani}, G. and {D'Odorico}, V. and {Dadi}, K. and {de Castro Le{\~a}o}, I. and {De Cia}, A. and {De Medeiros}, J.~R. and {Debras}, F. and {Debus}, M. and {Delorme}, A. and {Demangeon}, O. and {Derie}, F. and {Dessauges-Zavadsky}, M. and {Di Marcantonio}, P. and {Di Stefano}, S. and {Dionies}, F. and {Domiciano de Souza}, A. and {Doyon}, R. and {Dunn}, J. and {Egner}, S. and {Ehrenreich}, D. and {Faria}, J.~P. and {Ferruzzi}, D. and {Feruglio}, C. and {Fisher}, M. and {Fontana}, A. and {Frank}, B.~S. and {Fuesslein}, C. and {Fumagalli}, M. and {Fusco}, T. and {Fynbo}, J. and {Gabella}, O. and {Gaessler}, W. and {Gallo}, E. and {Gao}, X. and {Genolet}, L. and {Genoni}, M. and {Giacobbe}, P. and {Giro}, E. and {Gon{\c{c}}alves}, R.~S. and {Gonzalez}, O.~A. and {Gonz{\'a}lez-Hern{\'a}ndez}, J.~I. and {Gouvret}, C. and {Gracia T{\'e}mich}, F. and {Haehnelt}, M.~G. and {Haniff}, C. and {Hatzes}, A. and {Helled}, R. and {Hoeijmakers}, H.~J. and {Hughes}, I. and {Huke}, P. and {Ivanisenko}, Y. and {J{\"a}rvinen}, A.~S. and {J{\"a}rvinen}, S.~P. and {Kaminski}, A. and {Kern}, J. and {Knoche}, J. and {Kordt}, A. and {Korhonen}, H. and {Korn}, A.~J. and {Kouach}, D. and {Kowzan}, G. and {Kreidberg}, L. and {Landoni}, M. and {Lanotte}, A.~A. and {Lavail}, A. and {Lavie}, B. and {Lee}, D. and {Lehmitz}, M. and {Li}, J. and {Li}, W. and {Liske}, J. and {Lovis}, C. and {Lucatello}, S. and {Lunney}, D. and {MacIntosh}, M.~J. and {Madhusudhan}, N. and {Magrini}, L. and {Maiolino}, R. and {Maldonado}, J. and {Malo}, L. and {Man}, A.~W.~S. and {Marquart}, T. and {Marques}, C.~M.~J. and {Marques}, E.~L. and {Martinez}, P. and {Martins}, A. and {Martins}, C.~J.~A.~P. and {Martins}, J.~H.~C. and {Maslowski}, P. and {Mason}, C. and {Mason}, E. and {McCracken}, R.~A. and {Melo e Sousa}, M.~A.~F. and {Mergo}, P. and {Micela}, G. and {Milakovi{\'c}}, D. and {Molli{\`e}re}, P. and {Monteiro}, M.~A. and {Montgomery}, D. and {Mordasini}, C. and {Morin}, J. and {Mucciarelli}, A. and {Murphy}, M.~T. and {N'Diaye}, M. and {Nardetto}, N. and {Neichel}, B. and {Neri}, N. and {Niedzielski}, A.~T. and {Niemczura}, E. and {Nisini}, B. and {Nortmann}, L. and {Noterdaeme}, P. and {Nunes}, N.~J. and {Oggioni}, L. and {Olchewsky}, F. and {Oliva}, E. and {{\"O}nel}, H. and {Origlia}, L. and {{\"O}stlin}, G. and {Ouellette}, N.~N. -Q. and {Pall{\'e}}, E. and {Papaderos}, P. and {Pariani}, G. and {Pasquini}, L.},
        title = "{ANDES, the high resolution spectrograph for the ELT: science goals, project overview, and future developments}",
    booktitle = {Ground-based and Airborne Instrumentation for Astronomy X},
         year = 2024,
       editor = {{Bryant}, Julia J. and {Motohara}, Kentaro and {Vernet}, Jo{\"e}l. R.~D.},
       series = {Society of Photo-Optical Instrumentation Engineers (SPIE) Conference Series},
       volume = {13096},
        month = jul,
          eid = {1309613},
        pages = {1309613},
          doi = {10.1117/12.3017966},
archivePrefix = {arXiv},
       eprint = {2407.14601},
 primaryClass = {astro-ph.IM},
       adsurl = {https://ui.adsabs.harvard.edu/abs/2024SPIE13096E..13M}
}

@article{Queloz_2001,
       author = {{Queloz}, D. and {Mayor}, M. and {Udry}, S. and {Burnet}, M. and {Carrier}, F. and {Eggenberger}, A. and {Naef}, D. and {Santos}, N. and {Pepe}, F. and {Rupprecht}, G. and {Avila}, G. and {Baeza}, F. and {Benz}, W. and {Bertaux}, J. -L. and {Bouchy}, F. and {Cavadore}, C. and {Delabre}, B. and {Eckert}, W. and {Fischer}, J. and {Fleury}, M. and {Gilliotte}, A. and {Goyak}, D. and {Guzman}, J.~C. and {Kohler}, D. and {Lacroix}, D. and {Lizon}, J. -L. and {Megevand}, D. and {Sivan}, J. -P. and {Sosnowska}, D. and {Weilenmann}, U.},
        title = "{From CORALIE to HARPS. The way towards 1 m s$^{-1}$ precision Doppler measurements}",
      journal = {The Messenger},
         year = 2001,
        month = sep,
       volume = {105},
        pages = {1-7},
       adsurl = {https://ui.adsabs.harvard.edu/abs/2001Msngr.105....1Q}
}

@INPROCEEDINGS{Vogt_1994,
       author = {{Vogt}, S.~S. and {Allen}, S.~L. and {Bigelow}, B.~C. and {Bresee}, L. and {Brown}, B. and {Cantrall}, T. and {Conrad}, A. and {Couture}, M. and {Delaney}, C. and {Epps}, H.~W. and {Hilyard}, D. and {Hilyard}, D.~F. and {Horn}, E. and {Jern}, N. and {Kanto}, D. and {Keane}, M.~J. and {Kibrick}, R.~I. and {Lewis}, J.~W. and {Osborne}, J. and {Pardeilhan}, G.~H. and {Pfister}, T. and {Ricketts}, T. and {Robinson}, L.~B. and {Stover}, R.~J. and {Tucker}, D. and {Ward}, J. and {Wei}, M.~Z.},
        title = "{HIRES: the high-resolution echelle spectrometer on the Keck 10-m Telescope}",
    booktitle = {Instrumentation in Astronomy VIII},
         year = 1994,
       editor = {{Crawford}, David L. and {Craine}, Eric R.},
       series = {Society of Photo-Optical Instrumentation Engineers (SPIE) Conference Series},
       volume = {2198},
        month = jun,
        pages = {362},
          doi = {10.1117/12.176725},
       adsurl = {https://ui.adsabs.harvard.edu/abs/1994SPIE.2198..362V}
}

@inproceedings{Diego_1990,
title = "Final tests and commissioning of the UCL Echelle Spectrograph",
author = "Francisco Diego and Andy Charalambous and Fish, {Adrian C.} and Walker, {David D.}",
year = "1990",
month = jul,
day = "1",
doi = "10.1117/12.19119",
language = "English",
isbn = "0819402796",
volume = "1235",
series = "Proceedings of SPIE - The International Society for Optical Engineering",
publisher = "SPIE",
pages = "562--576",
editor = "Crawford, {David L.}",
booktitle = "Instrumentation in Astronomy VII",
note = "SPIE Astronomical Telescopes and Instrumentation for the 21st Century ; Conference date: 11-02-1990 Through 16-02-1990",
}

@INPROCEEDINGS{Dekker_2000,
       author = {{Dekker}, Hans and {D'Odorico}, Sandro and {Kaufer}, Andreas and {Delabre}, Bernard and {Kotzlowski}, Heinz},
        title = "{Design, construction, and performance of UVES, the echelle spectrograph for the UT2 Kueyen Telescope at the ESO Paranal Observatory}",
    booktitle = {Optical and IR Telescope Instrumentation and Detectors},
         year = 2000,
       editor = {{Iye}, Masanori and {Moorwood}, Alan F.},
       series = {Society of Photo-Optical Instrumentation Engineers (SPIE) Conference Series},
       volume = {4008},
        month = aug,
        pages = {534-545},
          doi = {10.1117/12.395512},
       adsurl = {https://ui.adsabs.harvard.edu/abs/2000SPIE.4008..534D}
}

@ARTICLE{Mayor_2003,
       author = {{Mayor}, M. and {Pepe}, F. and {Queloz}, D. and {Bouchy}, F. and {Rupprecht}, G. and {Lo Curto}, G. and {Avila}, G. and {Benz}, W. and {Bertaux}, J. -L. and {Bonfils}, X. and {Dall}, Th. and {Dekker}, H. and {Delabre}, B. and {Eckert}, W. and {Fleury}, M. and {Gilliotte}, A. and {Gojak}, D. and {Guzman}, J.~C. and {Kohler}, D. and {Lizon}, J. -L. and {Longinotti}, A. and {Lovis}, C. and {Megevand}, D. and {Pasquini}, L. and {Reyes}, J. and {Sivan}, J. -P. and {Sosnowska}, D. and {Soto}, R. and {Udry}, S. and {van Kesteren}, A. and {Weber}, L. and {Weilenmann}, U.},
        title = "{Setting New Standards with HARPS}",
      journal = {The Messenger},
         year = 2003,
        month = dec,
       volume = {114},
        pages = {20-24},
       adsurl = {https://ui.adsabs.harvard.edu/abs/2003Msngr.114...20M}
}

@INPROCEEDINGS{Crane_2010,
       author = {{Crane}, Jeffrey D. and {Shectman}, Stephen A. and {Butler}, R. Paul and {Thompson}, Ian B. and {Birk}, Christoph and {Jones}, Patricio and {Burley}, Gregory S.},
        title = "{The Carnegie Planet Finder Spectrograph: integration and commissioning}",
    booktitle = {Ground-based and Airborne Instrumentation for Astronomy III},
         year = 2010,
       editor = {{McLean}, Ian S. and {Ramsay}, Suzanne K. and {Takami}, Hideki},
       series = {Society of Photo-Optical Instrumentation Engineers (SPIE) Conference Series},
       volume = {7735},
        month = jul,
          eid = {773553},
        pages = {773553},
          doi = {10.1117/12.857792},
       adsurl = {https://ui.adsabs.harvard.edu/abs/2010SPIE.7735E..53C}
}

@ARTICLE{Pepe_2011,
       author = {{Pepe}, F. and {Lovis}, C. and {S{\'e}gransan}, D. and {Benz}, W. and {Bouchy}, F. and {Dumusque}, X. and {Mayor}, M. and {Queloz}, D. and {Santos}, N.~C. and {Udry}, S.},
        title = "{The HARPS search for Earth-like planets in the habitable zone. I. Very low-mass planets around <ASTROBJ>HD 20794</ASTROBJ>, <ASTROBJ>HD 85512</ASTROBJ>, and <ASTROBJ>HD 192310</ASTROBJ>}",
      journal = {\aap},
         year = 2011,
        month = oct,
       volume = {534},
          eid = {A58},
        pages = {A58},
          doi = {10.1051/0004-6361/201117055},
archivePrefix = {arXiv},
       eprint = {1108.3447},
 primaryClass = {astro-ph.EP},
       adsurl = {https://ui.adsabs.harvard.edu/abs/2011A&A...534A..58P}
}

@ARTICLE{Pepe_2021,
       author = {{Pepe}, F. and {Cristiani}, S. and {Rebolo}, R. and {Santos}, N.~C. and {Dekker}, H. and {Cabral}, A. and {Di Marcantonio}, P. and {Figueira}, P. and {Lo Curto}, G. and {Lovis}, C. and {Mayor}, M. and {M{\'e}gevand}, D. and {Molaro}, P. and {Riva}, M. and {Zapatero Osorio}, M.~R. and {Amate}, M. and {Manescau}, A. and {Pasquini}, L. and {Zerbi}, F.~M. and {Adibekyan}, V. and {Abreu}, M. and {Affolter}, M. and {Alibert}, Y. and {Aliverti}, M. and {Allart}, R. and {Allende Prieto}, C. and {{\'A}lvarez}, D. and {Alves}, D. and {Avila}, G. and {Baldini}, V. and {Bandy}, T. and {Barros}, S.~C.~C. and {Benz}, W. and {Bianco}, A. and {Borsa}, F. and {Bourrier}, V. and {Bouchy}, F. and {Broeg}, C. and {Calderone}, G. and {Cirami}, R. and {Coelho}, J. and {Conconi}, P. and {Coretti}, I. and {Cumani}, C. and {Cupani}, G. and {D'Odorico}, V. and {Damasso}, M. and {Deiries}, S. and {Delabre}, B. and {Demangeon}, O.~D.~S. and {Dumusque}, X. and {Ehrenreich}, D. and {Faria}, J.~P. and {Fragoso}, A. and {Genolet}, L. and {Genoni}, M. and {G{\'e}nova Santos}, R. and {Gonz{\'a}lez Hern{\'a}ndez}, J.~I. and {Hughes}, I. and {Iwert}, O. and {Kerber}, F. and {Knudstrup}, J. and {Landoni}, M. and {Lavie}, B. and {Lillo-Box}, J. and {Lizon}, J. -L. and {Maire}, C. and {Martins}, C.~J.~A.~P. and {Mehner}, A. and {Micela}, G. and {Modigliani}, A. and {Monteiro}, M.~A. and {Monteiro}, M.~J.~P.~F.~G. and {Moschetti}, M. and {Murphy}, M.~T. and {Nunes}, N. and {Oggioni}, L. and {Oliveira}, A. and {Oshagh}, M. and {Pall{\'e}}, E. and {Pariani}, G. and {Poretti}, E. and {Rasilla}, J.~L. and {Rebord{\~a}o}, J. and {Redaelli}, E.~M. and {Santana Tschudi}, S. and {Santin}, P. and {Santos}, P. and {S{\'e}gransan}, D. and {Schmidt}, T.~M. and {Segovia}, A. and {Sosnowska}, D. and {Sozzetti}, A. and {Sousa}, S.~G. and {Span{\`o}}, P. and {Su{\'a}rez Mascare{\~n}o}, A. and {Tabernero}, H. and {Tenegi}, F. and {Udry}, S. and {Zanutta}, A.},
        title = "{ESPRESSO at VLT. On-sky performance and first results}",
      journal = {\aap},
         year = 2021,
        month = jan,
       volume = {645},
          eid = {A96},
        pages = {A96},
          doi = {10.1051/0004-6361/202038306},
archivePrefix = {arXiv},
       eprint = {2010.00316},
 primaryClass = {astro-ph.IM},
       adsurl = {https://ui.adsabs.harvard.edu/abs/2021A&A...645A..96P}
}

@ARTICLE{Lohner-Bottcher_2019,
       author = {{L{\"o}hner-B{\"o}ttcher}, J. and {Schmidt}, W. and {Schlichenmaier}, R. and {Steinmetz}, T. and {Holzwarth}, R.},
        title = "{Convective blueshifts in the solar atmosphere. III. High-accuracy observations of spectral lines in the visible}",
      journal = {\aap},
         year = 2019,
        month = apr,
       volume = {624},
          eid = {A57},
        pages = {A57},
          doi = {10.1051/0004-6361/201834925},
archivePrefix = {arXiv},
       eprint = {1901.07606},
 primaryClass = {astro-ph.SR},
       adsurl = {https://ui.adsabs.harvard.edu/abs/2019A&A...624A..57L}
}

@ARTICLE{Deming_2024,
       author = {{Deming}, Drake and {Llama}, Joe and {Fu}, Guangwei},
        title = "{Precise Radial Velocities Using Line Bisectors}",
      journal = {\aj},
         year = 2024,
        month = jan,
       volume = {167},
       number = {1},
          eid = {34},
        pages = {34},
          doi = {10.3847/1538-3881/ad109f},
archivePrefix = {arXiv},
       eprint = {2312.03068},
 primaryClass = {astro-ph.SR},
       adsurl = {https://ui.adsabs.harvard.edu/abs/2024AJ....167...34D}
}

@BOOK{Kurucz_1984,
       author = {{Kurucz}, Robert L. and {Furenlid}, Ingemar and {Brault}, James and {Testerman}, Larry},
       title = "{Solar flux atlas from 296 to 1300 nm}",
       publisher={National Solar Observatory, Sunspot, NM},
       year = 1984,
       adsurl = {https://ui.adsabs.harvard.edu/abs/1984sfat.book.....K}
      }

@article{McWilliam_2026,
doi = {10.3847/1538-3881/ae45fd},
url = {https://doi.org/10.3847/1538-3881/ae45fd},
year = {2026},
month = {mar},
publisher = {The American Astronomical Society},
volume = {171},
number = {4},
pages = {233},
author = {McWilliam, Naomi and de Beurs, Zoë L. and Vanderburg, Andrew and Viaña, Javier and Mortier, Annelies and Buchhave, Lars A. and Collier Cameron, Andrew and Cosentino, Rosario and Dumusque, Xavier and Ghedina, Adriano and Lakeland, Ben and Lodi, Marcello and López-Morales, Mercedes and Sasselov, Dimitar and Sozzetti, Alessandro},
title = {Identifying Exoplanets with Deep Learning. VI. Enhancing Neural Network Mitigation of Stellar Activity RV Signals with Additional Metrics},
journal = {The Astronomical Journal}
}

@INPROCEEDINGS{Ycas_2010,
       author = {{Ycas}, Gabriel G. and {Quinlan}, Franklyn and {Osterman}, Steven and {Nave}, Gillian and {Diddams}, Scott A.},
        title = "{An optical frequency comb for infrared spectrograph calibration}",
    booktitle = {Ground-based and Airborne Instrumentation for Astronomy III},
         year = 2010,
       editor = {{McLean}, Ian S. and {Ramsay}, Suzanne K. and {Takami}, Hideki},
       series = {Society of Photo-Optical Instrumentation Engineers (SPIE) Conference Series},
       volume = {7735},
        month = jul,
          eid = {77352R},
        pages = {77352R},
          doi = {10.1117/12.857462},
       adsurl = {https://ui.adsabs.harvard.edu/abs/2010SPIE.7735E..2RY}
}

@article{Reiners_2024,
	author = {{Reiners, A.} and {Debus, M.} and {Sch\"afer, S.} and {Tiemann, E.} and {Zechmeister, M.}},
	title = {Accurate calibration spectra for precision radial velocities - Iodine absorption referenced by a laser frequency comb},
	DOI= "10.1051/0004-6361/202451389",
	url= "https://doi.org/10.1051/0004-6361/202451389",
	journal = {A\&A},
	year = 2024,
	volume = 690,
	pages = "A210",
}

@misc{keck_status,
  year=2026,
  title={Keck Instrument Status},
  key={KPF Status Summary 2026-02-01, },
  url={https://www2.keck.hawaii.edu/inst/kpf/status/},
  urldate={2026-05-21},
  note={Accessed: 2026-05-21}
  }

@misc{keck_stability,
  year=2026,
  
  key={KPF Semester 26A Stability Announcement 2026-02-01},
  title={Keck Semester 26A Stability Announcement},
  url={https://www2.keck.hawaii.edu/inst/kpf/announcements/2026-02-01_26A-status/},
  note={Accessed: 2026-05-21}
}

@misc{kpf_stab_announce,
  year=2025,
  key={KPF Stability Statement 2025-08-20},
  title={KPF Stabilitye Statement},
  url={https://www2.keck.hawaii.edu/inst/kpf/KPF%20Stability%20Statement%20-%20August%2015%202025.pdf},
  note={Accessed: 2026-05-21}
}
\bibliographystyle{aasjournal}



\end{document}